\documentclass[aps,showpacs,preprint,amsmath,amssymb,groupedaddress]{revtex4}

\usepackage{verbatim}
\usepackage[dvips]{graphicx}\unitlength=1mm
\def\slashed#1{\kern+0.1em /\kern-0.55em #1}

\newcommand{\Minv}[1]{M^{-1}_{#1}}
\renewcommand{\d}{\mathrm{d}}
\newcommand{\dM}[1]{\frac{\d M_{#1}}{\d \mu_{#1}}}
\newcommand{\ddM}[1]{\frac{\d^2 M_{#1}}{\d \mu^2_{#1}}}
\newcommand{\tr}{\text{tr}}

\begin{document}

\preprint{}

\author{Michael Cheng\footnote{Current address: Lawrence Livermore National Laboratory, \
Livermore, CA, 94550}}
\affiliation{
Department of Physics, Columbia University, New York, NY 10027, USA}

\author{Norman H. Christ}
\affiliation{
Department of Physics, Columbia University, New York, NY 10027, USA}

\author{Prasad Hegde}
\affiliation{
Department of Physics and Astronomy, SUNY, Stony Brook NY 11794-3800, USA}

\author{Frithjof Karsch}
\affiliation{
Physics Department, Brookhaven National Laboratory, Upton, NY 11973, USA}
\affiliation{
Fakult\"at f\"ur Physik, Universit\"at Bielefeld, D-33615 Bielefeld, Germany}

\author{Min Li}
\affiliation{
Department of Physics, Columbia University, New York, NY 10027, USA}

\author{Meifeng Lin\footnote{Current address: Department of Physics, Sloane Laboratory, Yale University, New Haven, CT 06520}}
\affiliation{Center for Theoretical Physics, 
Massachusetts Institute of Technology, Cambridge, MA  02139, USA}

\author{Robert D. Mawhinney}
\affiliation{
Department of Physics, Columbia University, New York, NY 10027, USA}

\author{Dwight Renfrew}
\affiliation{
Department of Physics, Columbia University, New York, NY 10027, USA}

\author{Pavlos Vranas}
\affiliation{
Lawrence Livermore National Laboratory, Livermore, CA, 94550, USA}

\title{The finite temperature QCD using 2+1 flavors of domain wall fermions at $\boldmath{N_t = 8}$}

\begin{abstract}
We study the region of the QCD phase transition using 2+1 flavors of 
domain wall fermions (DWF) and a $16^3 \times 8$ lattice volume with 
a fifth dimension of $L_s = 32$.  The disconnected light quark chiral 
susceptibility, quark number susceptibility and the Polyakov loop 
suggest a chiral and deconfining crossover transition lying between 155 and 185 MeV for our 
choice of quark mass and lattice spacing.  In this region the lattice 
scale deduced from the Sommer parameter $r_0$ is $a^{-1} \approx 1.3$ GeV, 
the pion mass is $\approx 300$ MeV and the kaon mass is approximately 
physical.  The peak in the chiral susceptibility implies a pseudo 
critical temperature $T_c = 171(10)(17)$ MeV where the first error is 
associated with determining the peak location and the second with our 
unphysical light quark mass and non-zero lattice spacing.  The effects 
of residual chiral symmetry breaking on the chiral condensate and 
disconnected chiral susceptibility are studied using several values 
of the valence $L_s$.
\end{abstract}

\pacs{11.15.Ha, 12.38.Gc, 11.30.Rd}

\date{November 10, 2009}

\maketitle

\section{Introduction}
The properties of strongly-interacting matter change dramatically
as the temperature is increased.  At a sufficiently
high temperature, the basic constituents of matter (quarks and gluons)
are no longer confined inside hadronic bound states, but exist
as a strongly interacting quark-gluon plasma (QGP).  The properties of the
QGP have been subject to significant theoretical and experimental study.
The physics of the transition region controls the initial formation of 
the QGP in a heavy-ion collision, as well as the details of hadronic 
freeze-out as the QGP expands and cools.  Thus, the transition
temperature and the order of the transition are of fundamental importance
in their own right and of particular interest to both the theoretical 
and experimental heavy-ion community.

The location and nature of the QCD phase transition has been extensively 
studied using lattice techniques with several different fermion actions 
\cite{Bazavov:2009zn,Cheng:2006qk,Bernard:2004je,Aoki:2006br,Chen:2000zu,
Bornyakov:2009qh}.  Recently, the most detailed studies of the transition 
temperature have been performed with different variants of the staggered 
fermion action~\cite{Bazavov:2009zn,Cheng:2006qk,Bernard:2004je,Aoki:2006br}.  
Although staggered fermions are computationally inexpensive, they suffer 
the disadvantage that they do not preserve the full SU(2)$\times$SU(2) 
chiral symmetry of continuum QCD, but only a $U(1)$ subgroup.  This lack 
of chiral symmetry is immediately apparent in the pion spectrum for 
staggered quarks, where there is only a single pseudo-Goldstone pion, 
while the other pions acquire additional mass from $O(a^2)$ flavor 
mixing terms in the action.

Thus, it is important to study the QCD phase transition using a 
different fermion discretization scheme.  The Wilson fermion
formulation is fundamentally different from the staggered approach and would be 
an obvious basis for an alternative approach.  However, Wilson fermions 
may be a poor alternative because in that formulation chiral symmetry 
is completely broken at the lattice scale and only restored in the continuum 
limit, the same limit in which the breaking of SU(2)$\times$SU(2) chiral symmetry 
in the staggered fermion formulation disappears.  

A particularly attractive fermion formulation to employ is that of domain
wall fermions~\cite{Kaplan:1992bt, Shamir:1993zy, Furman:1994ky}.  This is 
a variant of Wilson fermions in which a fifth dimension is introduced (the 
$s$ direction).  In this scheme, left and right-handed chiral states are 
bound to the four dimensional boundaries of the five-dimensional volume.  
The finite separation, $L_s$ between the left- and right-hand boundaries 
or walls allows some mixing between these left- and right-handed modes
giving rise to a residual chiral symmetry breaking.  However, in contrast
to Wilson fermions, this residual chiral symmetry breaking can be strongly 
suppressed by taking the fifth-dimensional extent ($L_s$) to be large.  

To leading order in an expansion in lattice spacing, the residual chiral 
symmetry breaking can be characterized by a single parameter, the residual
mass $m_\mathrm{res}$, which acts as an additive shift to the bare input 
quark mass.  Thus, the full continuum SU(2)$\times$SU(2) chiral symmetry 
can be reproduced to arbitrary accuracy by choosing $L_s$ sufficiently 
large, even at finite lattice spacing.  However, this good
control of chiral symmetry breaking comes with an approximate factor of 
$L_s$ increase in computational cost.

For these reasons, one of the first applications of the domain wall 
fermion approach was to the study of QCD thermodynamics using lattices
with a time extent of $N_t = 4$ and 6~\cite{Chen:2000zu}.  These early
results were quite encouraging, showing a clear signal for a physical,
finite temperature transition.   However, these were two-flavor 
calculations limited to quarks with relatively heavy masses on the 
order of that of the strange quark and with such large lattice spacings 
that higher order residual chiral symmetry breaking effects, beyond 
$m_\mathrm{res} \ne 0$, may have been important.

Given the substantial increase in computer capability and the deeper 
understanding of domain wall fermions that has been achieved over the 
past decade, it is natural to return to this approach.  Now significantly
smaller quark masses and much finer lattices with $N_t=8$ can be 
studied and important aspects of residual chiral symmetry breaking
can be recognized and explored.

This paper presents such a first study of the QCD finite temperature 
transition region using domain wall fermions at $N_t = 8$ and is 
organized as follows.  Section~\ref{sec:details} gives the details of 
our simulation, with regard to the choice of actions, simulation 
parameters, and algorithms.  Section~\ref{sec:finite} presents our 
results for finite-temperature observables such as the chiral 
condensate, chiral susceptibility, quark number susceptibility, 
Polyakov loop, and Polyakov loop susceptibility.  Section~\ref{sec:zero} 
gives results for the zero-temperature observables: the static 
quark potential and the hadron spectrum, that were calculated to 
determine the lattice spacing and quark masses in physical units.  
Section~\ref{sec:mres} discusses the effects of residual chiral symmetry 
breaking on our calculation and consistency checks of this finite 
temperature application of the domain wall method.  Section~\ref{sec:Tc} 
makes an estimate of the pseudo critical temperature $T_c$ which 
characterizes the critical region and its associated systematic errors.  
Finally, Section~\ref{sec:conclusion} presents our conclusions and
outlook for the future.

\section{Simulation Details}
\label{sec:details}

For our study we utilize the standard domain wall fermion action and 
the Iwasaki gauge action.  The properties of this combination of actions 
has been extensively studied at zero temperature by the RBC-UKQCD 
collaboration~\cite{Antonio:2006px, Allton:2007hx, Bernard:2000gd, Allton:2008pn}.

Using the data from Ref.~\cite{Antonio:2006px,Allton:2007hx, Antonio:2008zz, Allton:2008pn}, we 
extrapolated to stronger coupling in order to estimate the bare input 
parameters: the gauge coupling, input light quark mass, and input 
strange quark mass ($\beta$, $m_l$, $m_s$), appropriate for 
the region of the finite-temperature transition at $N_t=8$.  The
value of the critical gauge coupling was estimated to be $\beta_c \sim 2.00$ 
and the corresponding residual mass $m_\mathrm{res} \sim 0.008$ for 
$L_s = 32$.  As a result, we have used $m_l = 0.003$ and $m_s = 0.037$ 
for the input light and strange quark masses in all of our runs.  
This corresponds to $(m_l+m_\mathrm{res})/(m_s + m_\mathrm{res}) \approx 0.25$.

For the finite temperature ensembles, we have used a lattice volume of 
$16^3 \times 8$, with $L_s = 32$.  Table \ref{tab:parameters} shows the 
different values of $\beta$ that we chose, as well as the total number of 
molecular dynamics trajectories generated for each $\beta$.  In the 
immediate vicinity of the transition, we have approximately $2000-3000$ 
trajectories, with fewer trajectories as we move further away from the critical 
gauge coupling, $\beta_c$.
	
\begin{figure}[ht]
\begin{center}
\includegraphics[width=\textwidth]{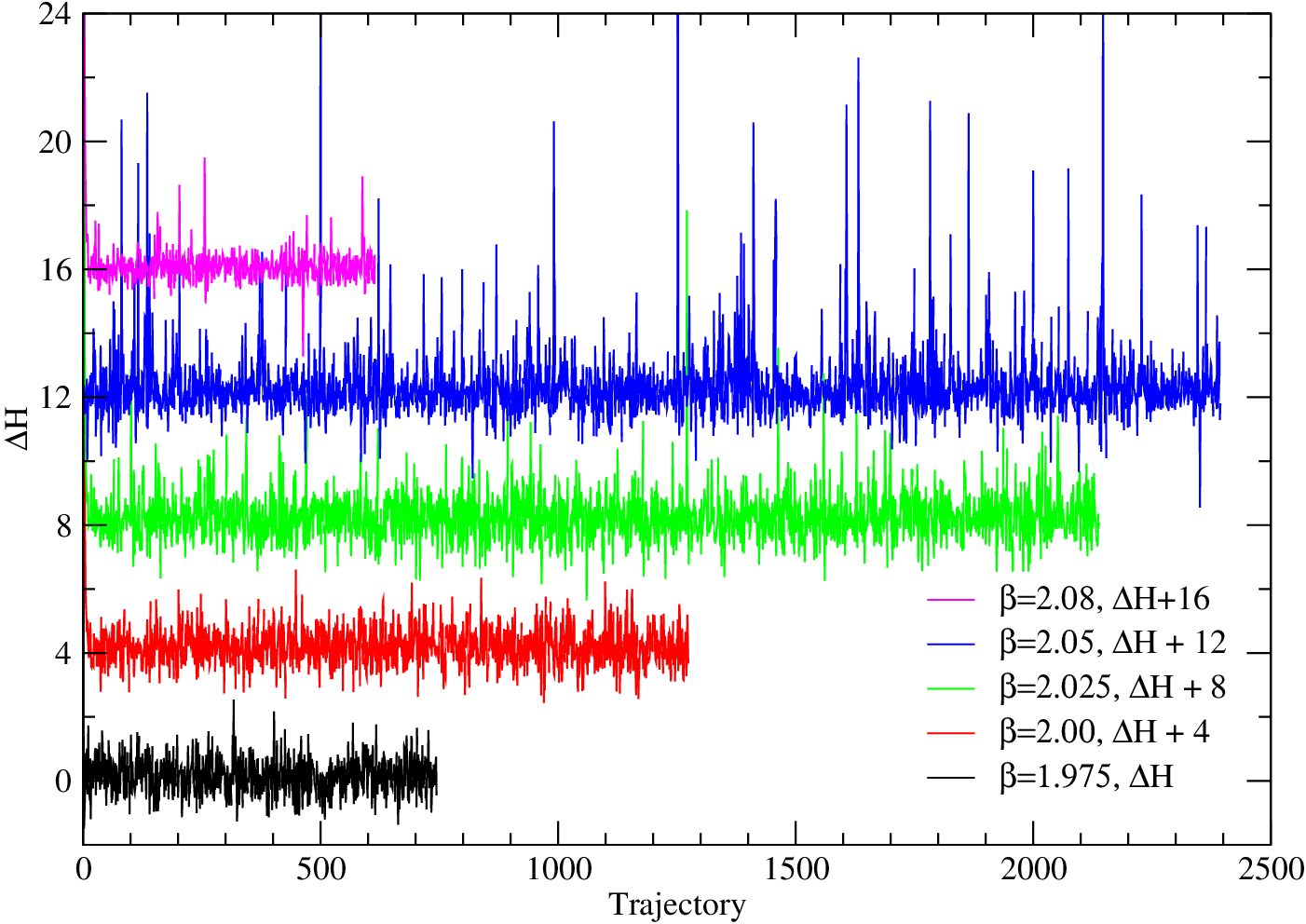}
\end{center}
\caption{The time history of $\Delta H$ for selected values of $\beta$. 
There is a vertical offset of 4 units between successive data sets with 
the lowest data set unshifted.}
\label{fig:hmc}
\end{figure}

We use the rational hybrid Monte Carlo (RHMC) algorithm 
\cite{Clark:2004cp,Clark:2006fx} to generate the dynamical field 
configurations.  An Omelyan integrator~\cite{deForcrand:2006pv,Takaishi:2005tz} 
with $\lambda = 0.22$ was used to numerically integrate the molecular 
dynamics trajectory.  A three-level integration scheme was used, where the 
force from the gauge fields was integrated with the finest time-step.  The 
ratio of the determinant of three flavors of strange quark to the determinant 
of three flavors of Pauli-Villars bosons was included at the intermediate 
time-step, while the ratio of the determinant of the two light quarks and 
the determinant of two strange pseudoquarks was integrated with the largest 
step-size.  The molecular dynamics trajectories were of unit length 
($\tau = 1$), with a largest step size of $\delta \tau = 0.2$ or 
$\delta \tau = 0.167$.   This allowed us to achieve an acceptance rate of 
approximately $75\%$.  Table~\ref{tab:parameters} summarizes the parameters 
that we have used for the finite temperature ensembles, as well as important 
characteristics of the RHMC evolution.  Figure~\ref{fig:hmc} shows the time 
history for $\Delta \mathcal{H}$ at a few selected gauge couplings.

We also generated 1200 trajectories at $\beta = 2.025$ with a volume of 
$16^3 \times 32$ and $L_s = 32$, also with $m_l = 0.003$ and $m_s = 0.037$.
We used these zero temperature configurations to determine the meson spectrum, 
as well as the static quark potential.

\begin{table}[bt]
\begin{center}
\begin{tabular}{c|ccccccccccc}
\hline
\hline
$\beta$ & 1.95 & 1.975 & 2.00 & 2.0125 & 2.025 & 2.0375 & 2.05 & 2.0625 & 2.08 & 2.11 & 2.14\\
\hline
Trajectories & ~745 & 1100 & 1275 & 2150 & 2210 & 2690 & 3015 & 2105 & 1655 & 440 & 490\\
Acceptance Rate & ~0.778 & 0.769 & 0.760 & 0.776 & 0.745 & 0.746 & 0.754 & 0.753 & 0.852 & 0.875 & 0.859\\
$\sqrt{\langle\Delta \mathcal{H}^2\rangle}$ & ~0.603 & 0.583 & 0.647 & 0.687 & 0.824 & 1.072 & 1.248 & 1.599 & 0.478 & 0.472 & 0.345\\
$\langle\exp(-\Delta \mathcal{H})\rangle$ & ~1.026 & 1.022 & 0.969 & 1.017 & 0.987 & 0.995 & 0.987 & 1.051 & 1.002 & 1.010 & 0.9979\\
\end{tabular}
\caption{Values for $\beta$, numbers of trajectories accumulated, results for the rms shift
in the RHMC Hamiltonian and the average exponentiated Hamiltonian shift (which should be unity). 
All runs were carried out with a trajectory length of 1 and an outer step size of 0.2 except 
for the case of $\beta=2.08$ where $\delta \tau = 0.167$ was used.}
\label{tab:parameters}
\end{center}
\end{table}

\section{Finite temperature observables}
\label{sec:finite}

For QCD with massless quarks, there is a true phase transition
from a low-temperature phase with spontaneous chiral symmetry breaking to a 
high temperature phase where chiral symmetry is restored.  If the quarks
have a finite mass ($m_f$), that explicitly breaks chiral symmetry,
the existence of a chiral phase transition persists for masses up to a critical
quark mass, $m_f < m_f^\mathrm{crit}$, above which the theory undergoes a smooth
crossover rather than a singular phase transition as the temperature is varied.  
The value of $m_f^\mathrm{crit}$ is poorly known and depends sensitively on 
the number of light quark flavors.  For a transition region dominated by two 
light quark flavors $m_f^\mathrm{crit}$ is expected to vanish and the transition 
to be second order only for massless quarks.  For three or more light flavors 
a first order region $0 \le m_f < m_f^\mathrm{crit}$ should be present.

\subsection{Chiral condensate}
\label{sec:chiral_condensate}

The order parameter that best describes the chiral phase transition is the
chiral condensate, $\langle\overline{\psi}_q\psi_q\rangle$, which vanishes in the
symmetric phase, but attains a non-zero expectation value in the 
chirally broken phase.  For quark masses above $m_f^\mathrm{crit}$, the chiral 
condensate will show only analytic behavior, but both the light and 
strange quark chiral condensates, $\langle\overline{\psi}_l \psi_l\rangle, 
\langle\overline{\psi}_s \psi_s\rangle$, and the disconnected part of their chiral 
susceptibilities, $\chi_l, \chi_s$, still contain information about the 
chiral properties of the theory in the vicinity of the crossover transition.
The chiral condensate and the disconnected chiral susceptibility for a single quark
flavor are defined as:
\begin{eqnarray}
\frac{\langle\overline{\psi}_q \psi_q\rangle}{T^3} & = & \frac{1}{VT^2}\frac{\partial \ln Z}{\partial m_q} = \frac{N_t^2}{N_s^3}\langle\mathrm{Tr}M_q^{-1}\rangle\\
\frac{\chi_q}{T^2} & = & \frac{1}{VT} \langle\left(\mathrm{Tr}M_q^{-1}\right)^2 - \langle \mathrm{Tr}M_q^{-1}\rangle^2\rangle = V T^3 \langle\left(\overline{\psi}_q\psi_q\right)^2 - \langle\overline{\psi}_q\psi_q\rangle^2\rangle
\end{eqnarray}
where $m_q$ is the mass of the single quark $q$ being examined, $T$ the temperature, 
$V$ the spatial volume and $N_t$ and $N_s$ are the number of lattice sites in the temporal
and spatial directions, respectively.

On our finite temperature ensembles, we calculate both the light
($m_l = 0.003$) and strange ($m_s = 0.037$) chiral condensates using 5
stochastic sources to estimate $\langle\overline{\psi}_q\psi_q\rangle$ on 
every fifth trajectory.  Using multiple stochastic sources on a given 
configuration allows us to extract an unbiased estimate of the fluctuations
in $\overline{\psi}_q\psi_q$ and to calculate the disconnected chiral 
susceptibility.  The Polyakov loop is calculated after every trajectory.

\begin{figure}[ht]
\begin{center}
\includegraphics[width=\textwidth]{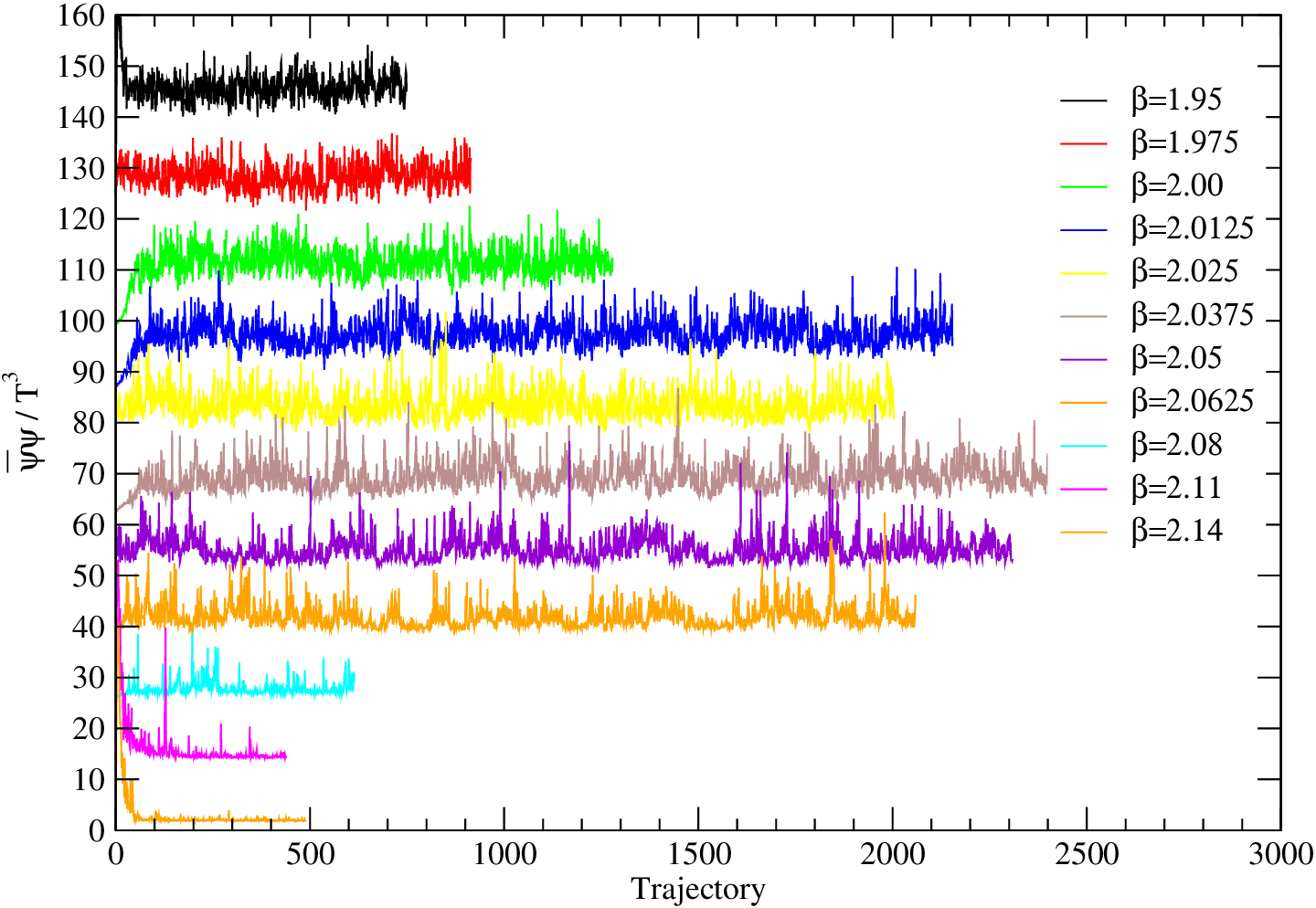}
\end{center}
\caption{The time history of $\overline{\psi}_l\psi_l$ for the light quarks.  There 
is a vertical offset of approximately 12 units between successive data sets with
the lowest set unshifted.}
\label{fig:pbp_history}
\end{figure}

Figures \ref{fig:pbp} and \ref{fig:pbp_sus} show the chiral condensate and
the disconnected part of the chiral susceptibility, respectively.  Examining 
the light and strange quark chiral condensates, it is difficult to precisely determine 
an inflection point.  Such an inflection point could be used to locate the 
mid-point of a thermal crossover.  We can also study the disconnected 
chiral susceptibility.  This is computed from the fluctuations in the chiral 
condensate and will show a peak near the location of the inflection point of the
chiral condensate.  Examining the time history of $\overline{\psi}_l\psi_l$ 
shown in Fig.~\ref{fig:pbp_history}, one can see that the fluctuations 
have a strong $\beta$ dependence.   We will identify the peak in these 
fluctuations with the location of the chiral crossover.  The chiral 
susceptibility shown in Fig. \ref{fig:pbp_sus}, has a clear peak near 
$\beta = 2.0375$.

At finite quark mass the chiral condensate contains an unphysical,
quadratically divergent, additive contribution coming from eigenvectors
of the Dirac operator with eigenvalue $\lambda \sim 1/a$.  These 
perturbative $\propto m_f/a^2$ terms will show no finite temperature
effects and obscure the physically important contribution from vacuum 
chiral symmetry breaking.  Since these terms enter both the light and 
strange condensates $\langle \overline{\psi}_l\psi_l\rangle$ and 
$\langle \overline{\psi}_s\psi_s\rangle$ in the same way it is appealing 
to remove this unphysical portion of $\langle \overline{\psi}_l\psi_l\rangle$ 
by subtracting $(m_l/m_s)\langle \overline{\psi}_s\psi_s\rangle$ from 
it~\cite{Cheng:2007jq}.  This should effectively remove the $m_l/a^2$ term 
from $\langle \overline{\psi}_l\psi_l\rangle$ while having little effect 
on the contribution from vacuum chiral symmetry breaking.  The result for
such a subtracted light chiral condensate is shown in Fig.~\ref{fig:pbp_sub}.

The exact form for this subtraction is complicated for domain wall fermions 
by the presence of residual chiral symmetry breaking.  In particular, the
factor $m_l/m_s$ might be constructed from the bare input quark masses
or from the more physical combination $(m_l+m_\mathrm{res})/(m_s+m_\mathrm{res})$.
As is discussed in Section~\ref{sec:pbp_res}, theoretical expectations and
our numerical results suggest that the short-distance, $1/a^2$ portion of
the chiral condensate will not show the $1/L_s$ behavior seen in the
residual mass so this latter subtraction would not be appropriate.
Instead, $\langle \overline{\psi}_q\psi_q\rangle$ approaches a constant
rapidly with increasing $L_s$ and in the limit of infinite $L_s$ the
ratio of the explicit chiral symmetry breaking parameters $m_l/m_s$ is 
the correct factor to use.  Thus, it is this approach which is shown in 
Fig.~\ref{fig:pbp_sub}.

\subsection{Polyakov loop}

For a pure $SU(3)$ gauge theory, there exists a first-order deconfining 
phase transition.  The relevant order parameter in this case is the Polyakov 
loop, $L$, which is related to the free energy of an isolated, static quark, 
$V_{hq}$: $L \sim \exp(-V_{hq}/T)$.  In the confined phase, producing an isolated 
quark requires infinite energy and the Polyakov loop vanishes.  However, at 
sufficiently high temperatures, the system becomes deconfined and the Polyakov 
loop acquires a non-vanishing expectation value in a sufficiently large volume.  
The Polyakov loop and its susceptibility are defined in terms of lattice variables
as:
\begin{eqnarray}
L& = & \frac{1}{3 N_s^3}\sum_{\vec r} 
   \mathrm{Tr} \left(\prod_{t=0}^{N_t-1} U_0(\vec r, t)\right)\\
\chi_L & = & N_s^3\left\{ \langle L^2\rangle - \langle L\rangle^2\right\}.
\end{eqnarray}

\begin{figure}[ht]
\begin{center}
\includegraphics[width=\textwidth]{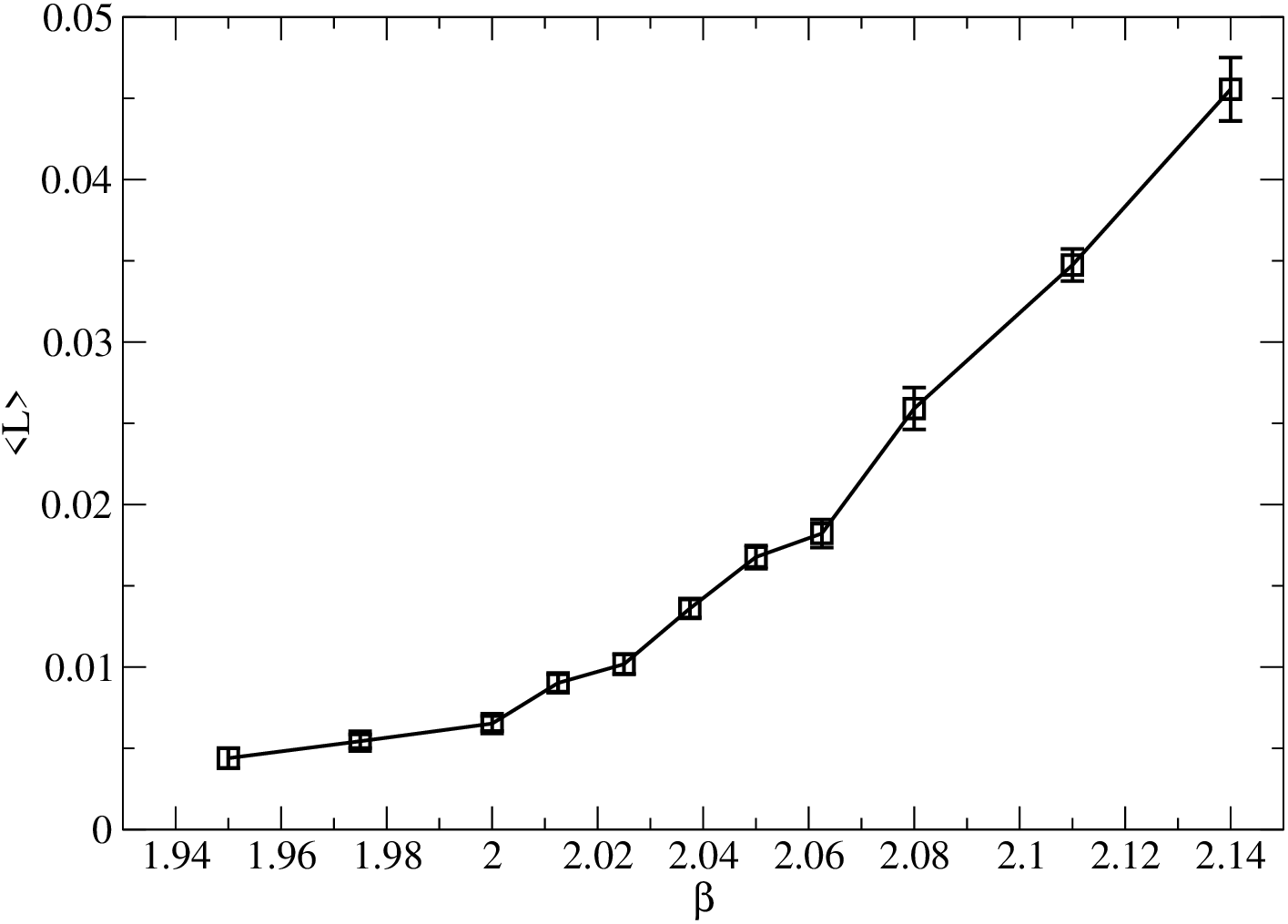}
\end{center}
\caption{Values obtained for the Polyakov loop as a
function of $\beta$.}
\label{fig:wline}
\end{figure}

\begin{figure}[ht]
\begin{center}
\includegraphics[width=\textwidth]{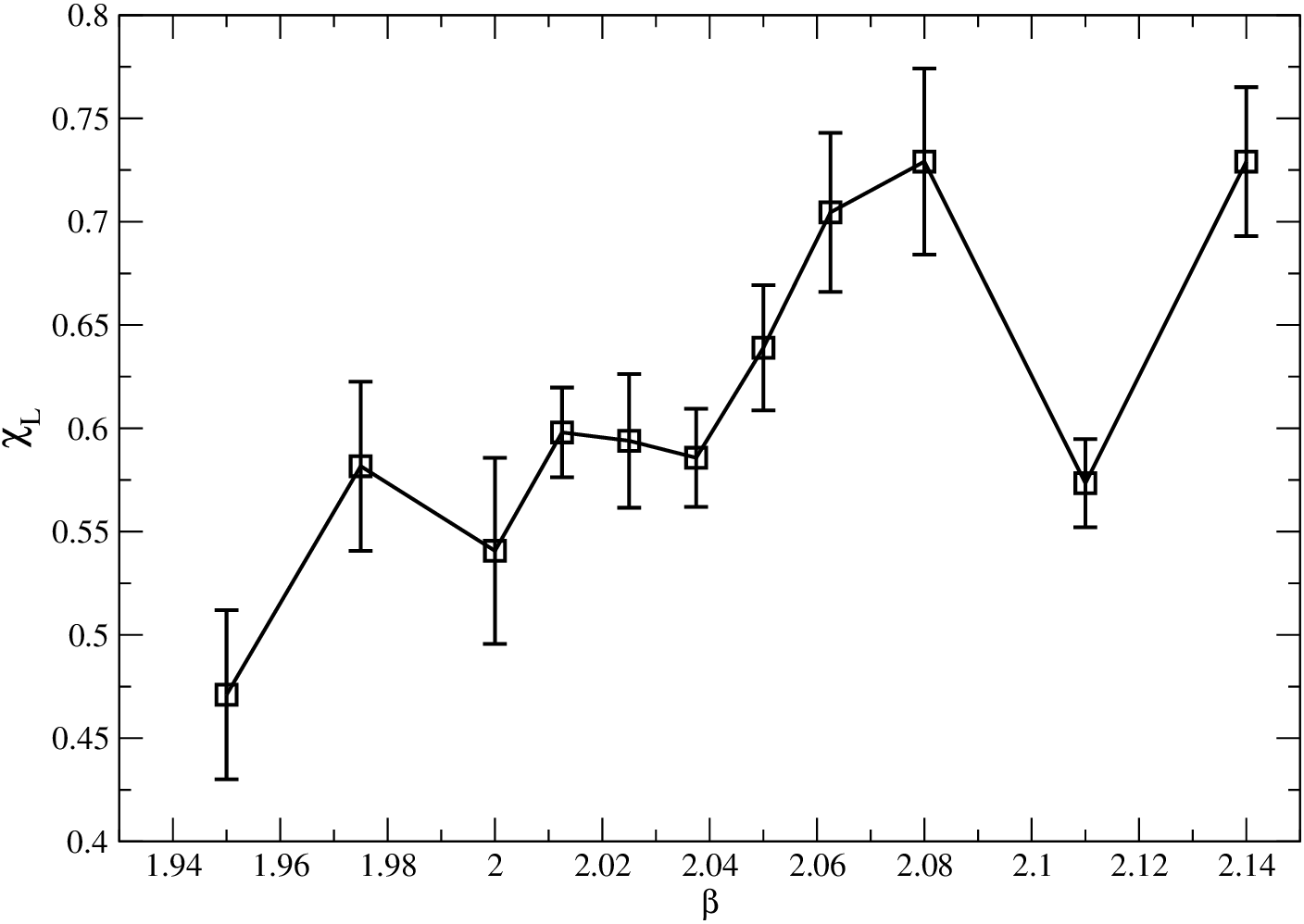}
\end{center}
\caption{The Polyakov loop susceptibility plotted as a function of $\beta$.}
\label{fig:wline_sus}
\end{figure}

Figures \ref{fig:wline} and \ref{fig:wline_sus} show the Polyakov loop
and the Polyakov loop susceptibility.  As in the case of the chiral condensate,
it is difficult to precisely locate an inflection point in the $\beta$
dependence of the Polyakov loop although the region where the Polyakov loop 
begins to increase more rapidly is roughly coincident with the peak in 
chiral susceptibility.  There is no well-resolved peak in the data for 
the Polyakov loop susceptibility, so we are unable to use this observable 
to locate the crossover region.  We list our results for these finite 
temperature quantities in Table~\ref{tab:finite_temp_obvs}.

\begin{table}
\begin{tabular}{r@{.}l|cc|cc|cc}
\hline
\hline
\multicolumn{2}{c|}{$\beta$} & $\langle\overline{\psi}_l\psi_l\rangle/T^3 $ & $\chi_l/T^2 $ & $\langle\overline{\psi}_s\psi_s\rangle/T^3$ & $\chi_s/T^2$ & $\langle L\rangle ~ (10^{-3})$ & $\chi_L$\\
\hline
1&95   & 22.8(2) & 6.4(17) & 40.9(1) & 3.5(8) & 4.40(62) & 0.47(4)\\
1&975  & 17.9(2) & 8.2(14) & 36.8(1) & 4.1(7) & 5.44(42) & 0.58(4)\\
2&00   & 13.5(2) & 9.4(27) & 33.2(1) & 2.7(7) & 6.52(47) & 0.54(5)\\
2&0125 & 11.6(2) & 16.4(20) & 31.6 & 5.7(7) & 9.02(53) & 0.60(2)\\
2&025  & 9.9(2) & 17.8(26) & 30.2(1) & 4.7(6) & 10.18(61)& 0.59(3)\\
2&0375 & 8.2(2) & 28.2(25) & 28.9(1) & 5.3(5) & 13.61(55)& 0.59(2)\\
2&05   & 6.0(2) & 20.5(18) & 27.4(1) & 4.5(8) & 16.77(71)& 0.64(3)\\
2&0625 & 5.1(2) & 20.7(27) & 26.6(1) & 4.2(5) & 18.22(86)& 0.70(4)\\
2&08   & 3.5(2) & 11.4(20) & 25.2(1) & 3.0(6) & 25.91(129)& 0.73(5)\\
2&11   & 2.37(7) & 3.7(30) & 23.51(5) & 0.9(2) & 34.74(99) & 0.57(2)\\
2&14   & 2.03(2) & 0.15(2) & 22.59(7) & 0.6(3) & 45.6(20) & 0.73(4)\\
\end{tabular}
\caption{Results obtained for the light and strange quark chiral
condensates and disconnected chiral susceptibilities as well as
the Polyakov loop and its susceptibility.}
\label{tab:finite_temp_obvs}
\end{table}

\subsection{Quark Number Susceptibilities}

Calculations performed with staggered and Wilson fermions at finite temperature have
shown that the analysis of thermal fluctuations of conserved charges, \textit{e.g.} baryon
number, strangeness or electric charge, gives sensitive information about the 
deconfining features of the QCD transition at high temperature. Charge fluctuations
are small at low temperature, rapidly rise in the transition region and approach the 
ideal gas Stefan-Boltzmann limit at high temperature. These generic features are easy 
to understand. Charge fluctuations
are small at low temperatures as charges are carried by rather heavy hadrons, while they
are large at high temperature where the conserved charges are carried by almost
massless quarks. Charge fluctuations therefore reflect deconfining aspects of the
QCD transition.

Thermal fluctuations of conserved charges can be calculated from diagonal and 
off-diagonal quark number susceptibilities which are defined as second derivatives of 
the QCD partition function with respect to quark chemical potentials~\cite{Gottlieb:1987ac}, 
($\mu_u,\ \mu_d,\ \mu_s$),
\begin{eqnarray}
\frac{\chi_2^f}{T^2} = \frac{2 c_2^f}{T^2} &=& \left. \frac{1}{VT^3}\frac{\partial^2
\ln Z(V,T,\mu_u,\mu_d,\mu_s)}{\partial (\mu_f/T)^2} \right|_{\mu_f=0}\nonumber\\
&=& \frac{1}{VT^3}\left\{\Bigg\langle\tr\left(\Minv{f}\ddM{f}\right)\Bigg\rangle -
\Bigg\langle\tr\left(\Minv{f}\dM{f}\Minv{f}\dM{f}\right) \Bigg\rangle \right.\nonumber \\
&&\;\;\;\;\;\;\;\;\; \left.      + 
\Bigg\langle \tr^2 \left(\Minv{f}\dM{f}\right)  \Bigg\rangle \right\}
\;\;, \;\; f=u,\ d,\ s \ ,
\label{fluc_diag}  \\
\frac{\chi_{11}^{fg}}{T^2} = \frac{c_{11}^{fg}}{T^2} &=& \left. \frac{1}{VT^3}\frac{\partial^2
\ln Z(V,T,\mu_u,\mu_d,\mu_s)}{\partial\mu_f/T\ \partial\mu_g/T} \right|_{\mu_g=\mu_f=0}
\nonumber \\
&=& \frac{1}{VT^3} \Bigg\langle\tr\left(\Minv{f}\dM{f}\right)
\tr\left(\Minv{g}\dM{g}\right)\Bigg\rangle
\;\;, \;\; f,g=u,\ d,\ s \ ,\ f\ne g\ ,
\label{fluc_uds}
\end{eqnarray}
where $c_2^f$ and $c_{11}^{fg}$ are the second-order coefficients in a Taylor
expansion of $p/T^4$.

In the DWF formalism the introduction of quark chemical potentials is 
straightforward~\cite{Bloch:2007xi,Hegde:2008nx,Gavai:2008ea}.  It 
follows the same approach used in other fermion discretization schemes 
\cite{Hasenfratz:1983ba}, {\it i.e.} in the fermion determinant for 
quarks of flavor $f$ the parallel transporters in forward [backward] 
time direction are multiplied with exponential factors $\exp (\mu_fa)$ 
[$\exp (-\mu_fa)$], respectively~\footnote{For other implementations 
of a chemical potential see Ref.~\cite{Gavai:2008ea, Gavai:2009vb}}.  
Since these time direction parallel transporters couple to the fermion 
fields for all locations $0 \le s < L_s$ in the fifth dimension, 
fermionic charge is assigned in a consistent way throughout the fifth 
dimension.  Just as in the case of the fermionic 
action~\cite{Furman:1994ky,Vranas:1997da}, a precaution 
must be taken to ensure that unphysical, 5-dimensional modes do not begin 
to contribute as $L_s$ becomes large.  The contribution of individual 
5-dimension modes, not bound to the $s=0$ or $s=L_s-1$ walls, will vanish 
in the continuum limit.  However, for finite lattice spacing and large 
$L_s$ the number of these modes may be sufficient to distort physical 
quantities.  In our calculation this is avoided by adding an additional 
compensating Pauli-Villars pseudo-fermion field for each quark flavor.
Thus, the chemical potential $\mu_f$ for each quark flavor enters the
time parallel transporters for both the light quark and the 
corresponding Pauli-Villars pseudo-fermion carrying that flavor.  These 
Pauli-Villars fields have $m_f=1$ and therefore satisfy anti-periodic 
boundary conditions in the 5-dimension.  Thus, they contribute no 
``physical'' 4-dimensional surface states but act to cancel any 
possible bulk contributions $\propto L_s$ introduced by the domain 
wall quarks.

Introducing chemical potentials for conserved charges, \textit{e.g.} baryon number ($\mu_B$),
strangeness ($\mu_S$) and electric charge ($\mu_Q$), allows us to define susceptibilities 
(charge fluctuations)  by taking derivatives with respect to these chemical 
potentials~\footnote{Quark chemical potentials and chemical potentials for conserved
charges are related through $\displaystyle{\mu_u=\mu_B/3 + 2 \mu_Q/3}$,
$\displaystyle{\mu_d=\mu_B/3 - \mu_Q}/3$,
$\displaystyle{\mu_s=\mu_B/3 - \mu_Q/3 - \mu_S}$
(see for instance \cite{Cheng:2008zh})},
\begin{eqnarray}
\frac{\chi_2^X}{T^2} = \frac{2 c_2^X}{T^2} & = & \frac{1}{VT^3}\frac{\partial^2
\ln Z(V,T,\mu_B,\mu_S,\mu_Q)}{\partial (\mu_X/T)^2}|_{\mu_X=0}
\;\;, \;\; X=B,\ S,\ Q
\label{fluc}
\end{eqnarray}

Expressed in terms of quark number susceptibilities, one finds,
\begin{eqnarray}
c_2^S &=& c_2^s \\
c_2^B &=& \frac{1}{9}\left(2 c^u_2 + c^s_2 +  c^{ud}_{11} +
2 c^{us}_{11}\right)  \\
c_2^Q &=& \frac{1}{9}\left(5 c^u_2 + c^s_2 - 2 c^{ud}_{11} -
c^{us}_{11}\right)
\label{fluc_BQS}
\end{eqnarray}
Similar to the chiral susceptibility, the two derivatives appearing in Eq.~\ref{fluc_diag}
generate 'disconnected' and 'connected' contributions to the flavor diagonal susceptibilities. 
The mixed susceptibilities defined in Eq.~\ref{fluc_uds}, on the other hand, only receive
contributions from disconnected terms. As the disconnected terms are much more noisy than
the connected terms, those susceptibilities that are dominated by contributions from the
latter are generally easier to calculate.
This makes the electric charge susceptibility and the isospin susceptibility,
$\displaystyle{c_2^I = \left(2c^u_2 - c^{ud}_{11}\right)/4}$, most
suitable for our current, exploratory analysis with domain wall fermions.

\begin{table}
\begin{center}
\begin{tabular}{l|c|c|c|cccc}
\hline
\hline
~~$\beta$ & measurements & separation & random vectors  & $c_2^u$ & $c_2^s$ & $c_2^I$ & $c_2^Q$\\
\hline
1.95 & 73 & 10 & 200 & 0.08(11) & 0.01(5) & 0.046(8) & 0.060(10)\\
1.975 & 61 & 10 & 200 & 0.03(10) & 0.03(7) & 0.070(8) & 0.085(10)\\
2.0125~& 125 & 10 & 150 & 0.22(6) & 0.16(2) & 0.119(7) & 0.148(10)\\
2.025 & 71 & 20 & 150 & 0.30(5) & 0.19(3) & 0.141(6) & 0.176(8)\\
2.0375~& 96 & 20 & 150 & 0.30(6) & 0.16(2) & 0.160(6) & 0.205(8)\\
2.05 & 81 & 25 & 150 & 0.38(5) & 0.25(4) & 0.191(9) & 0.243(11)\\
2.0625~ & 111 & 10 & 150 & 0.32(6) & 0.24(4) & 0.200(9) & 0.252(10)\\
2.11 & 35 & 10 & 100 & 0.51(6) & 0.44(5) & 0.233(11) & 0.303(14)\\
2.14 & 40 & 10 & 100 &  0.51(3) & 0.43(2) & 0.256(4) & 0.333(5)\\
\end{tabular}
\end{center}
\caption{Details of the calculation of quark number susceptibilities.
The column labeled "measurements" gives the number of measurements that 
were performed.  That labeled "separation" gives the number of time 
units between those measurements while the "random vectors" column
gives the number of random vectors used in each measurement.}
\label{tab:sim_details}
\end{table}
Computing the susceptibilities involves measuring traces of operators. We used stochastic 
estimators with 100-200 random vectors per configuration. Our measurements are summarized in 
Table~\ref{tab:sim_details}. Some of the results presented here have been shown 
previously~\cite{Hegde:2008rm}.

\begin{figure}[ht]
\begin{center}
\includegraphics[width=\textwidth]{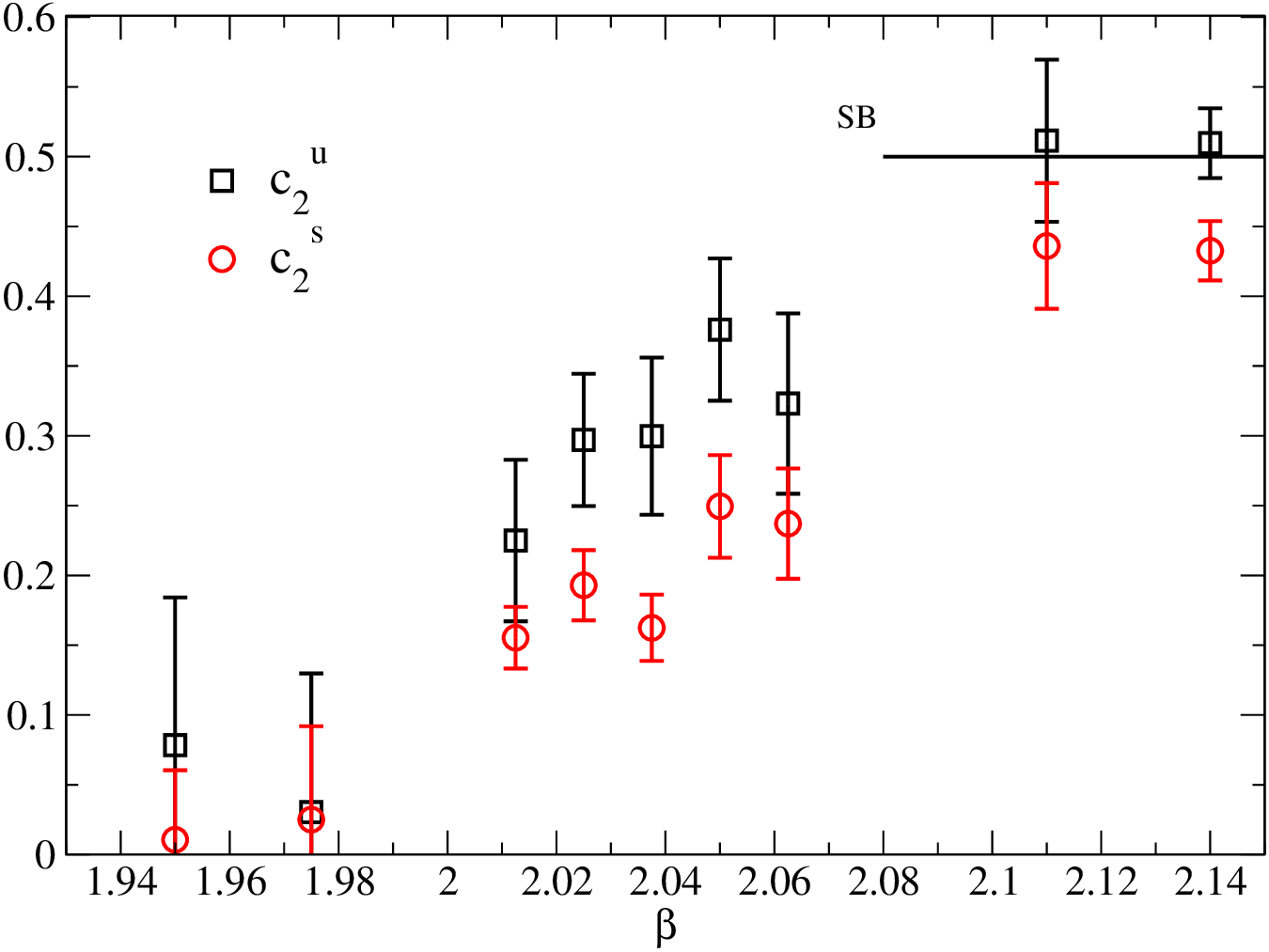}
\end{center}
\caption{The light and strange quark number susceptibilities $c_2^u$ and $c_2^s$
plotted as a function of $\beta$.}
\label{fig:results-1}
\end{figure}

In Fig.~\ref{fig:results-1}, we show our results for the diagonal, light and strange 
quark number, susceptibilities $c_2^{u}$ and $c_2^s$, respectively.  We see that these 
susceptibilities do transit from a low value to a high one as $\beta$ increases. However, 
given the current statistical accuracy of our calculation, it is difficult to assign any definite 
value of $\beta$ around which the transition takes place. To a large extent the fluctuations
observed in the data arise from contributions of off-diagonal susceptibilities,
$c_{11}^{fg}$, with $f\equiv g$. In fact, with our current limited statistics these 
susceptibilities vanish within errors and therefore only contribute noise to the 
diagonal susceptibilities.

\begin{figure}[!tbh]
\begin{center}
\begin{minipage}{0.75\textwidth}
\includegraphics[width=\textwidth]{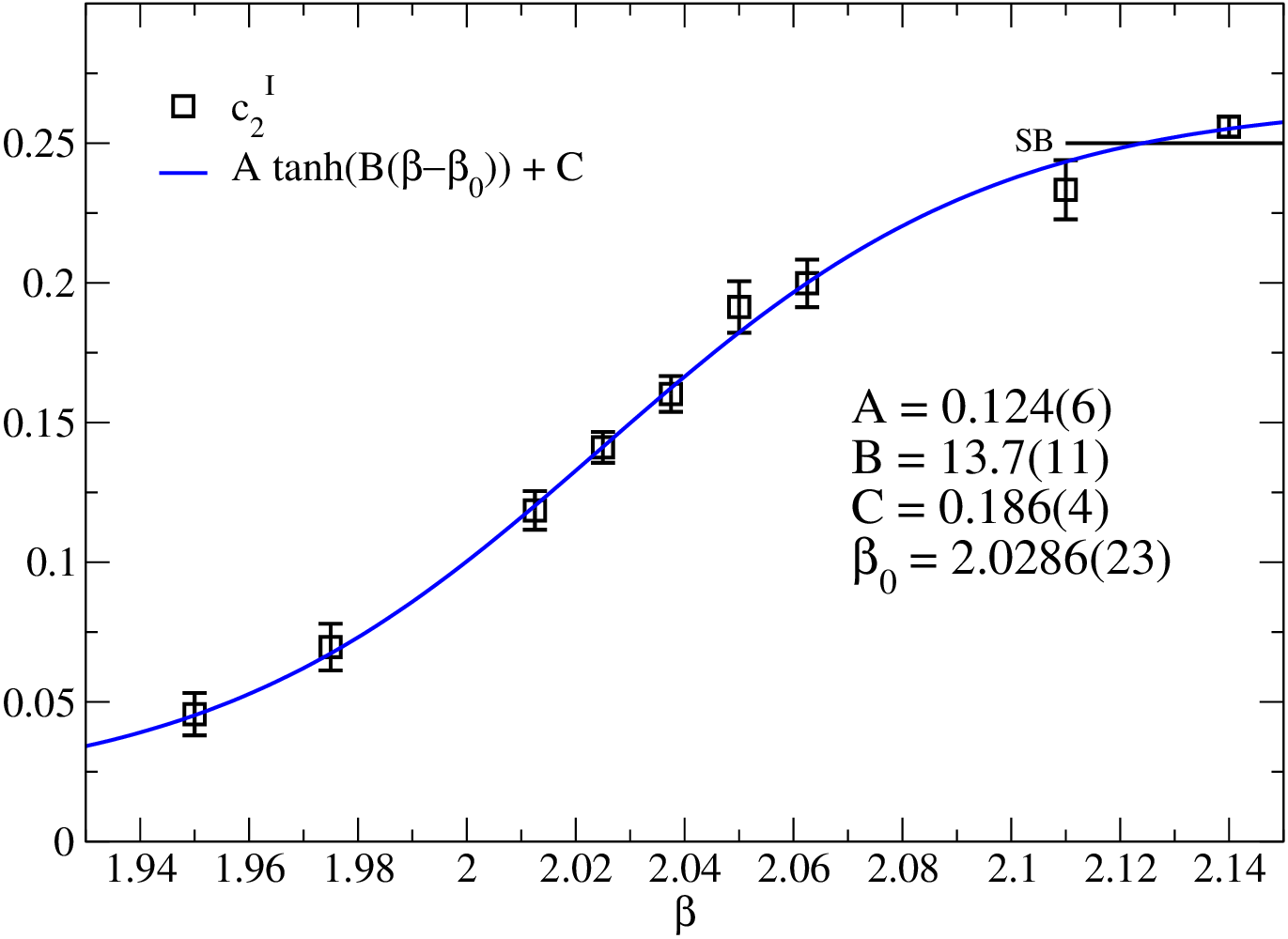}
\vspace{0.4cm}
\end{minipage}
\begin{minipage}{0.75\textwidth}
\includegraphics[width=\textwidth]{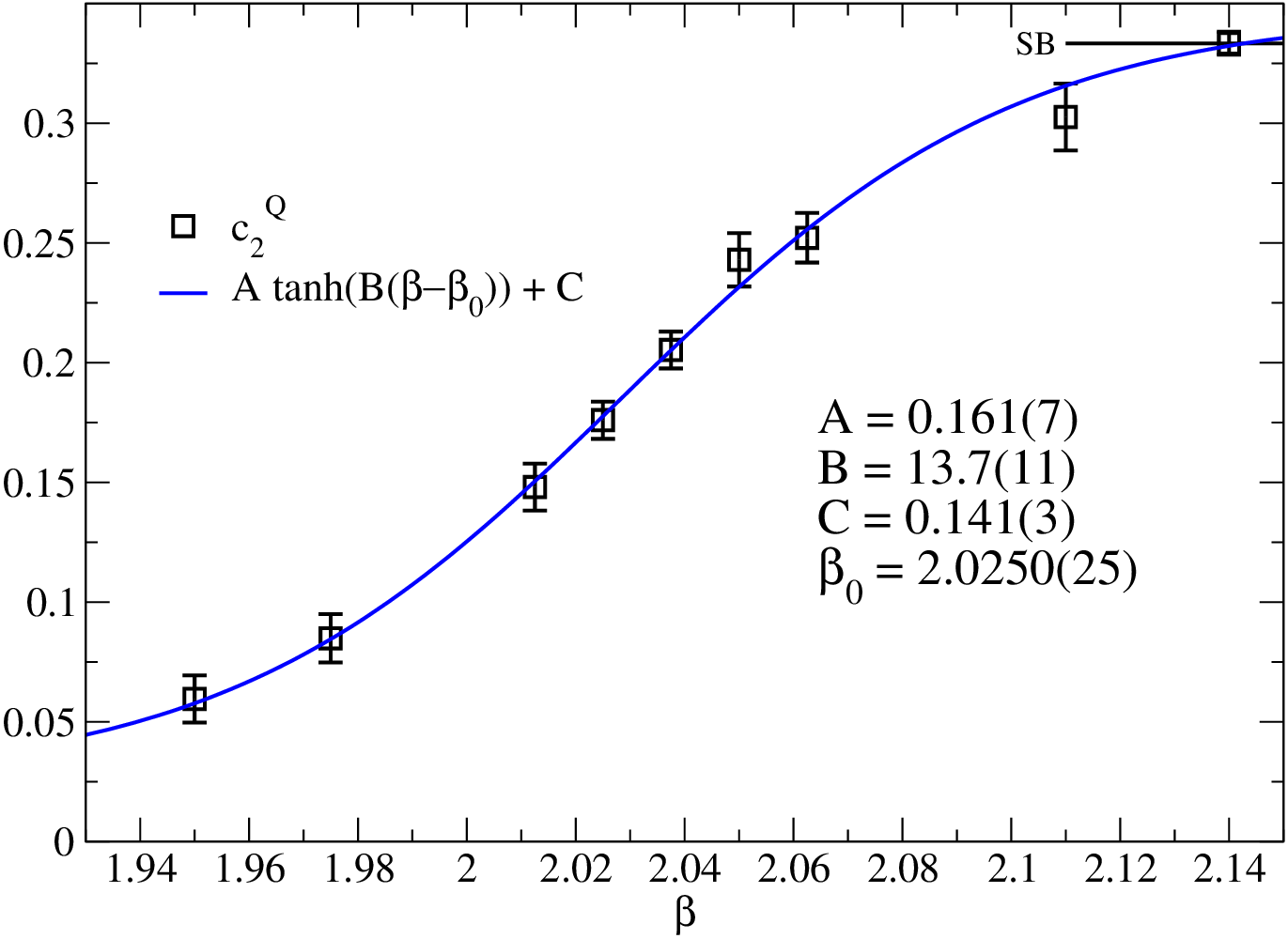}
\end{minipage}
\end{center}
\vspace{-0.3cm}
\caption{The susceptibilities $c_2^Q$ and $c_2^I$ plotted versus $\beta$. 
The lines show fits based on the hyperbolic ansatz, $f_I(\beta)$, given 
in Eq.~\ref{eq:fitab}.  The legend also gives the fit parameters, which includes
the location of the inflection point, $\beta_0$.}
\label{fig:results-2}
\end{figure}

The disconnected parts however, either completely or partially cancel out in the two 
susceptibilities $c_2^I$ and $c_2^Q$. As a result, one obtains much better results for 
these quantities, as seen in Fig.~\ref{fig:results-2}.   

We have tried to determine the inflection point for the electric charge and isospin
susceptibilities, which may serve as an estimate for the transition point, although
the slope of these observables also receives contributions from the regular part of 
the free energy. We have fit the data using two different fit ans\"atze,
\begin{eqnarray}
f_I(\beta) &=& A \tanh (B(\beta -\beta_0)) +C \; ,\nonumber \\
f_{II}(\beta) &=& A_3+B_3 \beta + C_3 \beta^2 +D_3 \beta^3 \; .
\label{eq:fitab}
\end{eqnarray} 
To estimate systematic errors in the fits we performed fits for the entire data set
as well as in  limited ranges by leaving out one or two data points at the lower
as well as upper edge of the $\beta$-range covered by our data sample. From this we
find inflection points in the range $2.024\le \beta_0\le 2.037$ for $c_2^I$ and 
$2.024\le \beta_0\le 2.034$ for $c_2^Q$. Summarizing this analysis we therefore 
conclude that the inflection points in the electric charge and isospin susceptibilities
coincide within statistical errors and are given by $\beta_0 = 2.030(7)$.  This
is in good agreement with the determination of a pseudo-critical coupling obtained
from the location of peak in the chiral susceptibility, $\beta = 2.0375$, found 
in Section~\ref{sec:chiral_condensate}.

\section{Zero temperature observables}
\label{sec:zero}

In this section we present the results for physical quantities at
zero temperature computed on a $16^3 \times 32$ lattice for $\beta=2.025$
which, as Fig.~\ref{fig:pbp_sus} suggests, lies in the lower temperature
part of the $N_t=8$ transition region.

\subsection{Static quark potential}
To determine the lattice scale, we measured the static quark-anti-quark 
correlation function, $W(r,t)$, on 148 configurations (every 5 MD 
trajectories from 300-1035) on these zero temperature configurations. 
The quantity $W(r,t)$ is the product of two spatially separated sequences
of temporal gauge links connecting spatial hyperplanes, each containing 
links that have been fixed to Coulomb gauge~\cite{Bernard:2000gd, Li:2006gra}:  
\begin{eqnarray}
W(r,t) &=& \frac{1}{N_\mathrm{pairs}(r)} \sum_{\left|\vec r_1 - \vec r_2\right|=r}
\mbox{tr}\Bigl\{ U_0(\vec r_1,0)  U_0(\vec r_1,1) \dots  U_0(\vec r_1,t-1) \\
&&\hskip 1.5in \cdot  U_0^\dagger(\vec r_2,t-1) \dots 
                      U_0^\dagger(\vec r_2,1) U_0^\dagger(\vec r_2,0)\Bigr\},
\nonumber
\end{eqnarray}
where $N_\mathrm{pairs}(r)$ is the number of pairs of lattice points with a given
spatial separation $r$.  In our calculation
the results obtained from orienting the ``time'' axis along each of the 
four possible directions are also averaged together.  The time dependence of 
$W(r,t)$ was then fit to an exponential form in order to extract the static 
quark potential $V(r)$:
\begin{equation}
W(r,t) = c(r) \exp\left(-V(r) t\right).
\end{equation}
The potential $V(r)$ was subsequently fit to the Cornell form, and used to 
determine the Sommer parameter $r_0$, as defined below:
\begin{eqnarray}
V(r) & = & -\frac{\alpha}{r} + \sigma r + V_0 \label{eq:cornell} \\
\left(r^2 \frac{d V(r)}{dr}\right)_{r=r_0} & = & 1.65\;.
\end{eqnarray}
Table~\ref{tab:r0} gives the details of the fit which determines the 
parameters $\alpha$ and $\sigma$ of Eq.~\ref{eq:cornell} and results in 
a value of $r_0/a = 3.08(9)$.  For the physical value of $r_0$, we use 
the current standard result $r_0 = 0.469(7)$ fm~\cite{Gray:2005ur}.  
This gives a lattice spacing $a \approx 0.15$ fm, or $a^{-1} \approx 1.3$ GeV.  
It should be emphasized that this value for $r_0$ has been determined 
for a single light quark mass and no extrapolation to the physical value of
the light quark mass has been performed.  This failure to extrapolate to
a physical value for the light quark mass is likely to result in an
overestimate of the lattice spacing $a$ by about 3\%.

\begin{table}
\begin{tabular}{cccccc}
\hline
\hline
$\beta$ & $r_0/a$ & $a^{-1}$ (GeV) & ~t fit range~ &~ r fit range~ & $\chi^2/$dof.\\
\hline
2.025 & 3.08(9) & 1.30(4) & $(4,9)$ & $(\sqrt{3}, 6)$ & 1.03\\
\hline
\end{tabular}
\caption{Results for $r_0$.  The errors are calculated by the jackknife method, with 
data binned into blocks, each containing 10 molecular dynamics time units.}
\label{tab:r0}
\end{table}

\subsection{Meson mass spectrum}
In addition to the static quark potential, we also calculated the meson spectrum 
on the same zero temperature ensemble at $\beta = 2.025$.  The meson spectrum
was determined using 55 configurations, separated by 10 MD time units, from 500 and 1040.
Table~\ref{tab:spectrum} gives the results for $m_{\rho}$ and $m_{\pi}$ for three
different valence mass combinations, as well as their values in the chiral limit from
linear extrapolation.    Equating the physical value of $m_{\rho} = 776$ MeV with the 
chirally extrapolated lattice value gives a lattice scale of $a^{-1} = 1.26(11)$ GeV, 
which is consistent with the scale determined from $r_0$.  Examining the data 
for the light pseudoscalar meson, we find $m_{\pi} \approx 308$ MeV, somewhat larger than 
twice the mass of the physical pion.  For the kaon, we have $m_K \approx 496$ MeV, 
very close to the physical kaon mass.

\begin{table}[hbt]
\centering
\begin{tabular}{cccccccc}
\hline
\hline
$m_{x}^{val}$ & $m_y^{val}$ & $m_\mathrm{avg}$ & fit range & $m_{\rho} a$ & $\chi^2$/dof & $m_{\pi} a$ & $\chi^2$/dof\\
\hline
0.003 &  0.003 & 0.0030 & 8-16 & 0.646(63) & 0.3(4) & 0.2373(20) & 2.4(11)\\
0.003 &  0.037 & 0.0200 & 8-16 & 0.716(23) & 0.8(7) & 0.3815(15) & 2.0(10)\\
0.037 &  0.037 & 0.0370 & 8-16 & 0.776(10) & 2.2(11)& 0.4846(11) & 1.2(8)\\
\hline
 &  $-m_\mathrm{res}$ & & & 0.617(56) &  & 0.073(6) &\\
\hline
\end{tabular}
\caption{The calculated masses $m_{\rho}$ and $m_{\pi}$ for various combinations
of valence quark mass.  The last line represents extrapolation of the light quark
mass to $m_\mathrm{avg} = (m_x+m_y)/2 = -m_\mathrm{res}$.}
\label{tab:spectrum}
\end{table}

\section{Residual Chiral Symmetry breaking}
\label{sec:mres}

We now examine the central question in such a coarse-lattice calculation
using domain wall fermions: the size and character of the residual chiral 
symmetry breaking effects.  We examine the residual mass computed at finite 
temperature, its $L_s$ dependence and the dependence of the chiral condensate
on $L_s$.  In both cases we examine the value of $L_s = 32$ used for the 
dynamical quarks as well as "non-unitary'', valence values of $L_s$ varying 
between 8 and 128.

\subsection{Residual Mass}

One of the primary difficulties with the calculation presented here is the 
rather large residual chiral symmetry breaking at the parameters that we 
employ.  This manifests itself in a value for the residual mass, 
$m_\mathrm{res}$ which is larger than the input light quark mass, 
$m_{ud} = 0.003$ over almost the entire temperature range of our calculation.

For the Iwasaki gauge action, the residual chiral symmetry breaking has
been extensively studied by the RBC-UKQCD collaboration for $\beta \ge 2.13$ 
and $L_s=16$~\cite{Antonio:2006px, Allton:2007hx, Antonio:2008zz, Allton:2008pn}.  
However, the lattice ensembles that we use here are significantly coarser, 
resulting in larger residual chiral symmetry breaking, even for our increased 
value of $L_s = 32$.

\begin{table}[hbt]
\begin{center}
\begin{tabular}{ccc}
\hline
\hline
$\beta$  & $m_\mathrm{res}$ ($m_f = 0.003$) & $m_\mathrm{res}$ ($m_f = 0.037$)\\
\hline
1.95 & 0.0253(5) & 0.0244(5)\\
2.00 & 0.0105(3) & 0.0095(2)\\
2.025& 0.0069(3) & 0.0059(3)\\
2.05 & 0.0046(5) & 0.0034(2)\\
2.08 & 0.0023(5) & 0.0016(2)\\
2.11 & 0.0011(2) & 0.0009(1)\\
2.14 & 0.0010(4) & 0.0006(2)\\
\hline
\end{tabular}
\caption{The residual mass as a function of $\beta$ computed on the
finite temperature, $16^3 \times 8$ lattice volume.}
\label{tab:mres_beta}
\end{center}
\end{table}

\begin{table}[hbt]
\begin{center}
\begin{tabular}{ccc}
\hline
\hline
$m_\textrm{val}$ & $m_{\rm res}$ & fit range\\
\hline
0.003 & 0.006647(84) & 8-16\\
0.020 & 0.006227(74) & 8-16\\
0.037 & 0.005835(71) & 8-16\\
\hline
0.000 & 0.006713(85) & \\
\hline
\end{tabular}
\caption{The residual mass as a function of valence quark mass computed on the
zero temperature, $16^3 \times 32$ lattice volume with $\beta = 2.025$, with the extrapolated
$m_\textrm{val} \rightarrow 0$ value.}
\label{tab:mres}
\end{center}
\end{table}

\begin{table}[hbt]
\centering
\begin{tabular}{ccc}
\hline
\hline
$L_s$ & $m_\mathrm{res}$ ($m_f = 0.003$) & $m_\mathrm{res}$ ($m_f = 0.037$)\\
\hline
8  & 0.0529(9) & 0.0508(7)\\
16 & 0.0235(5) & 0.0220(4)\\
32 & 0.0105(3) & 0.0095(2)\\
64 & 0.0048(3) & 0.0044(3)\\
128& 0.0024(2) & 0.0025(2)\\
\hline
\end{tabular}
\caption{The residual mass as a function of the valence $L_s$ computed on a 
$16^3 \times 8$ lattice volume with $\beta = 2.00$.}
\label{tab:mres_Ls}
\end{table}

Table \ref{tab:mres_beta} shows our results for $m_\mathrm{res}$ on several 
of the $16^3 \times 8$ finite temperature ensembles.  We follow the standard method,
described for example in Ref.~\cite{Antonio:2006px}, determining the residual mass 
by computing the ratio of the midpoint correlator to the pion correlator evaluated 
at source-sink separations sufficiently large to suppress short-distance lattice 
artifacts.  This is most easily done on these finite temperature lattices by 
choosing the source-sink separation to lie in a spatial rather than temporal 
direction.  

Table \ref{tab:mres} gives $m_\mathrm{res}$ on the $16^3 \times 32$ ensemble at 
$\beta = 2.025$ where the correlators are measured in the temporal direction.  
It is important to observe that the values of $m_\mathrm{res}$ determined at
$\beta=2.025$ at finite and zero temperature, 0.0069(5) and 0.006647(84) respectively, are
consistent.  This is an important check on the domain wall method since $m_\mathrm{res}$
should be a temperature-independent constant representing the leading long-distance 
effects of residual chiral symmetry breaking.

\begin{figure}[ht]
\begin{center}
\includegraphics[width=\textwidth]{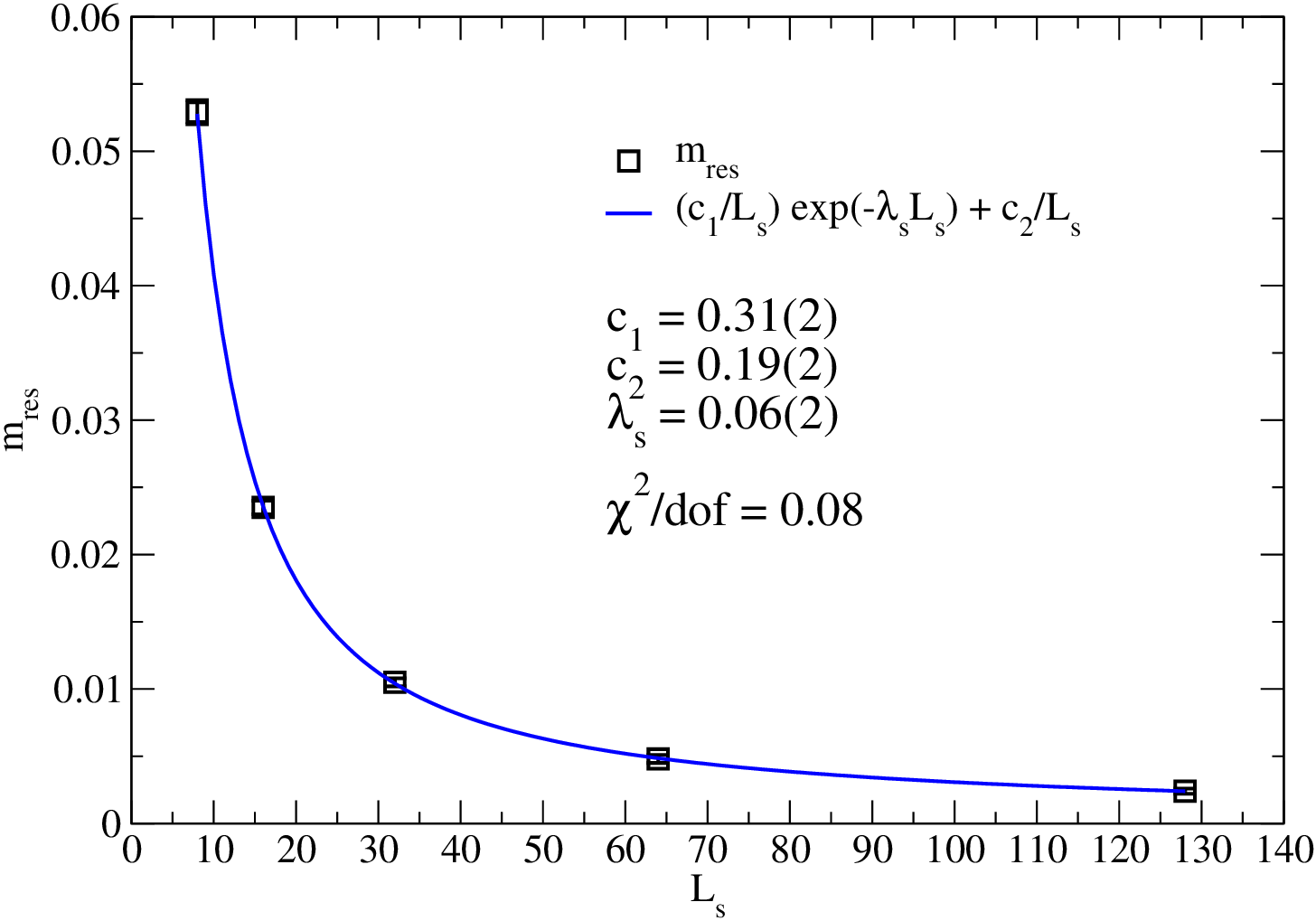}
\end{center}
\caption{The residual mass $m_{\mathrm{res}}$ is plotted versus $L_s$ for 
$\beta = 2.00$, $16^3\times8$.  The fit to Eq.~\ref{eqn:mres} is also shown.}
\label{fig:mres_b2.00}
\end{figure}

Table \ref{tab:mres_Ls} shows results for $m_\mathrm{res}$ evaluated at different 
values for the valence $L_s$ at $\beta = 2.00$.  The expected behavior of $m_\mathrm{res}$
as a function of $L_s$ is given by~\cite{Antonio:2008zz}:
\begin{equation}
m_\mathrm{res}(L_s) = \frac{c_1}{L_s}\exp(-\lambda_c L_s) + \frac{c_2}{L_s}.
\label{eqn:mres}
\end{equation}
Here the exponential term comes from extended states with eigenvalues near 
the mobility edge, $\lambda_c$, while the $1/L_s$ piece reflects the presence of 
localized modes with small eigenvalues and is proportional to the density of 
such small eigenvalues at $\lambda = 0$~\cite{Golterman:2003qe,Golterman:2004cy,
Golterman:2005fe,Antonio:2008zz}.  This formula describes our data very well
as can be seen from Fig.~\ref{fig:mres_b2.00} where both the data presented in
Table~\ref{tab:mres} and the resulting fit to Eq.~\ref{eqn:mres} are shown.
The proportionality of $m_\mathrm{res}$ to $1/L_s$ shown in Table~\ref{tab:mres_Ls} 
for $L_s \ge 32$ indicates that our choice of $L_s = 32$ has effectively 
suppressed the exponential term in Eq.~\ref{eqn:mres} but that a large contribution 
remains from the significant density of near-zero eigenvalues on our relatively 
coarse lattice.

Since we have chosen the input light quark mass $m_l = 0.003$ to be fixed
for the different values of $\beta$, the strong dependence of $m_\mathrm{res}$ 
on $\beta$ shown in Table~\ref{tab:mres_beta} means that the total light 
quark mass, $m_q = m_l + m_\mathrm{res}$, changes significantly in the 
crossover region, from $m_q \approx 0.0075$ at $\beta = 2.05$ increasing 
to $m_q \approx 0.013$ at $\beta = 2.00$.  This substantial increase may 
significantly affect the quantities whose temperature dependence we
are trying to determine.

\subsection{Chiral condensate and susceptibility at varying $L_s$}
\label{sec:pbp_res}

The change in the total quark mass as we vary $\beta$ is expected to cause
a distortion of the chiral susceptibility curve that we use to locate the 
crossover transition.  In order to understand how this varying mass affects 
our results, we have computed the chiral condensate and its susceptibility 
with different choices for the valence $L_s$ and valence $m_l$ at several 
values of $\beta$.
  
\begin{figure}[ht]
\begin{center}
\includegraphics[width=\textwidth]{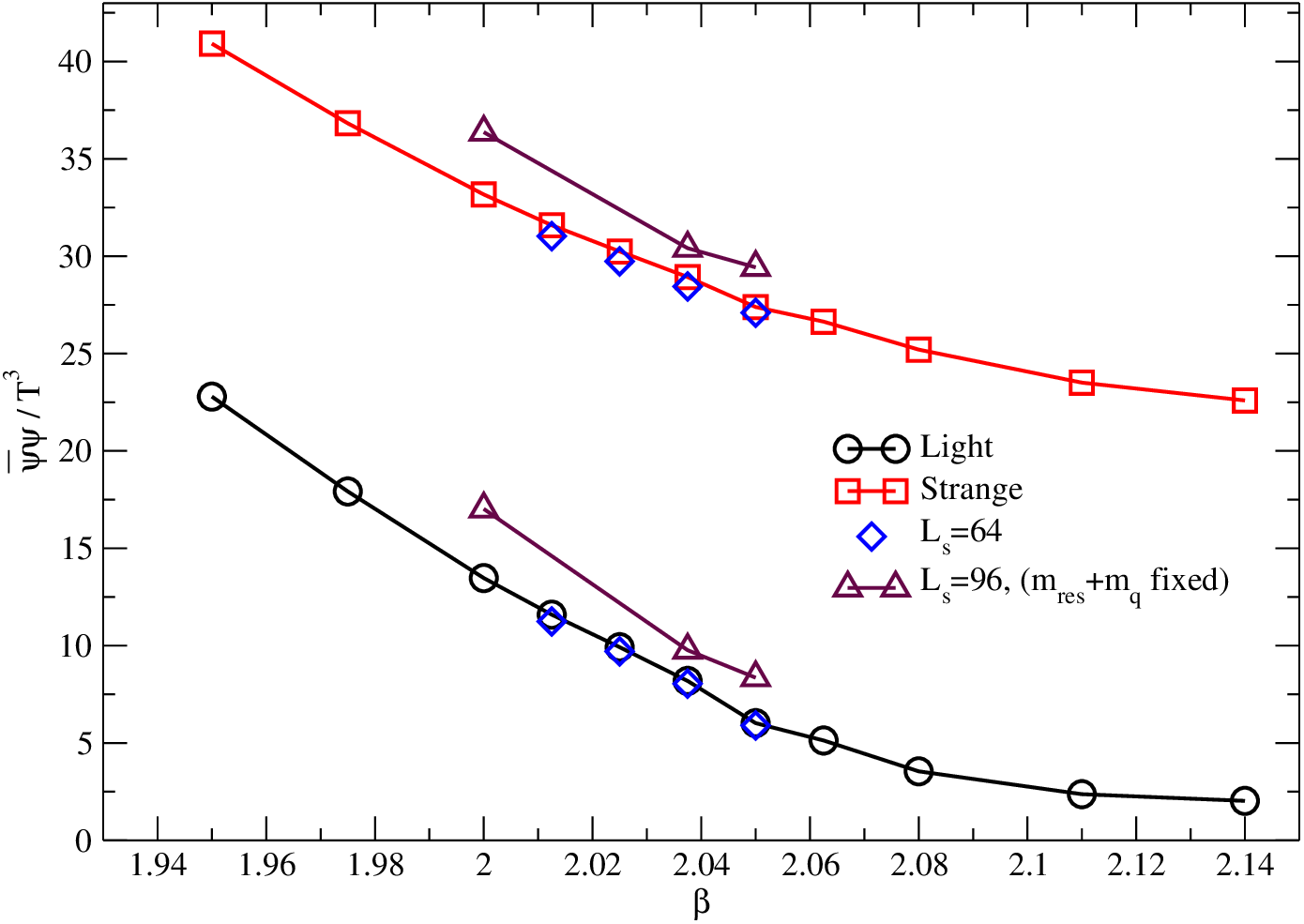}
\end{center}
\caption{Unitary values for $\langle\overline{\psi}_l\psi_l\rangle$ and 
$\langle\overline{\psi}_s\psi_s\rangle$ (the circles and squares respectively) 
for $L_s = 32$, as well as additional measurements with $L_s = 64$ and 
$L_s = 96$ for the valence quarks.  For the $L_s = 96$ measurements, $m_l$ 
and $m_s$ are adjusted so that values for the sum $m_q + m_{res}$ are 
approximately the same as those for $L_s = 32$.}
\label{fig:pbp}
\end{figure}

\begin{figure}[ht]
\begin{center}
\includegraphics[width=\textwidth]{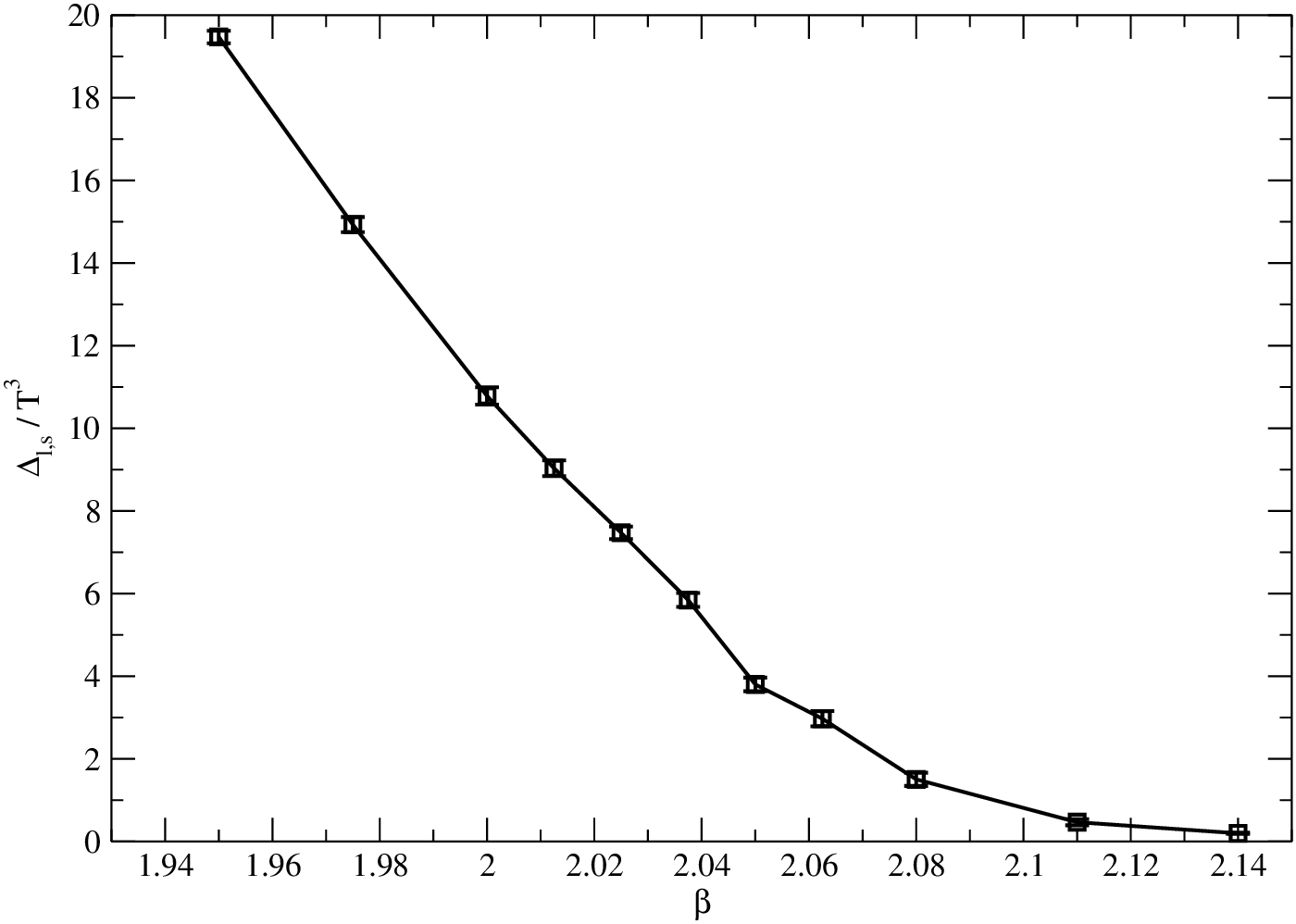}
\end{center}
\caption{The subtracted light-quark chiral condensate, $\Delta_{l,s}= 
\langle\overline{\psi}_l\psi_l\rangle - 
m_l/m_s\langle\overline{\psi}_s\psi_s\rangle$ as a function
of $\beta$.  This subtraction removes the uninteresting $m_l/a^2$
contribution from $\langle\overline{\psi}_l\psi_l\rangle$,
leaving a quantity which more accurately describes vacuum
chiral symmetry breaking.  This improvement is easily seen
for the larger values of $\beta$, above the transition region,
where this subtracted quantity vanishes, in contrast to the 
non-zero behavior seen for $\langle\overline{\psi}_l\psi_l\rangle$
in Fig.~\ref{fig:pbp}.}
\label{fig:pbp_sub}
\end{figure}

In one set of measurements, we increased $L_s$ from 32 to 64, while keeping
the input quark masses fixed at $m_l = 0.003$ and $m_s = 0.037$.  This has the 
result of reducing the total light and strange quark masses, as the residual
masses are reduced by approximately a factor of two.  In another set of
measurements, we increased $L_s$ to 96 but adjusted the input quark masses to
compensate for the reduced residual mass so that the total light and strange
quark masses, $m_l + m_{res}$ and $m_s + m_{res}$ respectively, matched those
in the $L_s = 32$ calculation for each value of beta.  Finally, for one value 
of the gauge coupling, $\beta = 2.0375$, we used several choices of valence 
$L_s$ (8, 16, 24, 48) at fixed input quark mass $(m_l,m_s) = (0.003, 0.037)$ 
in order to examine the $L_s$ dependence of our observables at fixed $\beta.$
Table \ref{tab:pbpLs} gives the results of these measurements.
Figures \ref{fig:pbp} and \ref{fig:pbp_sus} show the results with the
valence $L_s = 64$ and $L_s = 96$ in context with the $L_s = 32$ results.

\begin{table}[hbt]
\centering
\begin{tabular}{cc|ccc|ccc}
\hline
\hline
$L_s$ & $\beta$ & $m_l$ & $\langle\overline{\psi}_l\psi_l\rangle/T^3$ & $\chi_l/T^2$ & $m_s$ & $\langle\overline{\psi}_s\psi_s\rangle/T^3$ & $\chi_s/T^2$\\
\hline
8 & 2.0375 ~& ~0.003 & 26.6(1) & 7.2(8) & ~0.037 & 45.5(1) & 4.3(5)\\
16 & & & 10.8(1) & 12.4(1) & & 31.1(1) & 4.4(5)\\
24 & & & 8.6(1) & 17.8(2) & & 29.4(1) & 4.6(6)\\
48 & & & 7.8(2) & 33.2(5) & & 28.5(1) & 5.0(8)\\
\hline
64 & 2.0125 ~& ~0.003 &  11.2(2) & 32.3(3) & ~0.037 & 31.0(1) & 6.5(6)\\
& 2.025  & & 9.7(1) & 32.6(4) & & 29.7(1) & 5.1(8)\\
& 2.0375 & & 8.0(2) & 46.2(8) & & 28.4(1) & 4.9(7)\\
& 2.05   & & 5.9(2) & 39.0(4) & & 27.1(1) & 5.3(7)\\
\hline
96 & 2.00 ~& ~0.0078 & 17.0(4) & 13.6(36) & ~0.0418 & 36.4(2) & 1.9(11)\\
& 2.0375 ~& ~0.0063 & 9.8(1) & 24.8(26) & ~0.0403 & 30.4(1) & 4.9(6)\\
& 2.05 ~& ~0.0070 & 8.4(1) & 20.2(23) & ~0.0410 & 29.4(1) & 4.1(6)\\
\hline
\end{tabular}
\caption{Results for $\langle\overline{\psi}_q\psi_q\rangle$ and the 
corresponding disconnected susceptibility in which some of the values 
for $L_s$ and $m_l$, assigned to the quark loop present in the 
$\overline{\psi}_q\psi_q$ observable, differ from those that appear 
in the quark determinant.}
\label{tab:pbpLs}
\end{table}

\begin{figure}[ht]
\begin{center}
\includegraphics[width=\textwidth]{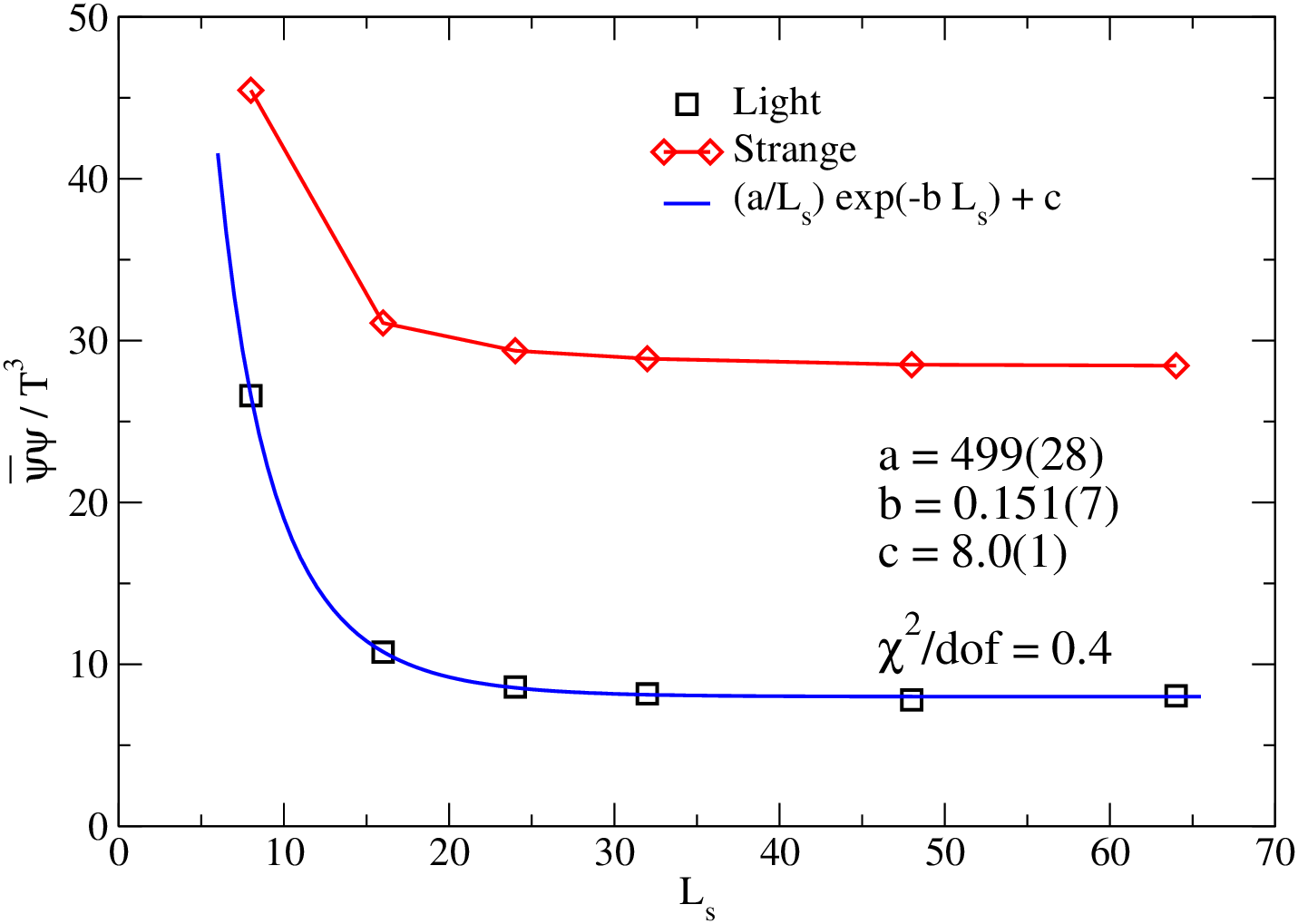}
\end{center}
\caption{Chiral condensate versus the valence $L_s$ for $\beta = 2.0375$, on a 
$16^3\times8$ lattice volume.  The fit to Eq.~\ref{eqn:condensate_fit} for $\overline{\psi}_l\psi_l/T^3$ is also shown.}
\label{fig:pbp_b2.0375}
\end{figure}

From Fig.~\ref{fig:pbp}, we see that increasing $L_s$ from 32 to 64
while keeping the input quark masses fixed does not have much effect on the 
chiral condensate for each $\beta$ at which we measure.  On the other hand, 
using $L_s = 96$ and larger input quark masses causes a noticeable increase
in the chiral condensate.  A closely related phenomenon can be found in
Fig. \ref{fig:pbp_b2.0375} which shows the dependence of 
$\langle\overline{\psi}_q\psi_q\rangle$ on $L_s$ at the single value of $\beta = 2.0375$.  
For small values of $L_s$, there is a strong $L_s$ dependence, but the chiral 
condensate quickly plateaus to an approximately constant value for $L_s > 32$, 
even though $m_\mathrm{res}$ and thus the total light quark mass is still 
changing significantly as $L_s$ increases above 32.

This contrast between the $L_s$ dependence of $\langle\overline{\psi}_q\psi_q\rangle$
and $m_{\rm res}$ can be made more precise if we attempt to fit the $L_s$
dependence of $\langle\overline{\psi}_q\psi_q\rangle$ by a single exponential, omitting
the power law piece that is important in $m_{\rm res}(L_s)$:
\begin{equation}
\langle\overline{\psi}_q\psi_q\rangle(L_s) = \frac{a}{L_s}~\exp(-b L_s)+c.
\label{eqn:condensate_fit}
\end{equation}
This fit describes the data very well, giving $\chi^2$/dof = 0.4, in strong
contrast to $m_{\rm res}(L_s)$ where the $c_2/L_s$ term in Eq.~\ref{eqn:mres}
is required to fit the data.  Thus, it appears that the contribution of
the localized modes, responsible for the $c_2/L_s$ term in Eq.~\ref{eqn:mres},
is much less important for the chiral condensate than for the residual
mass.  

In fact, this is to be expected.  The localized states are rather special.
They are associated with the near zero modes of the 4-D Wilson Dirac operator 
evaluated at a mass equal to the domain wall height, $-M_5$.  They are 
non-perturbative and appear when topology changes.  They are thus related 
to continuum physics and are limited in number.  In contrast, the extended 
states which give the exponential term $\exp(-\lambda_c L_s)/L_s$ can be seen in 
perturbation theory, correspond to large, $O(1/a)$ eigenvalues of 
$D_W^{4D}(-M_5)$ and are far more numerous with a density given by 
four-dimensional free-field phase space at the $\lambda \sim 1/a$ scale.  
Since the perturbative contribution to the dimension-one residual mass 
behaves as $1/a$ while that to the dimension-three chiral condensate as
$1/a^3$, it is to be expected that the non-perturbative, localized states 
will play a much larger role in the former.

If we accept that the $L_s$ behavior of the chiral condensate differs in
this way from that of the residual mass, then the behavior of the chiral
condensate shown in Fig.~\ref{fig:pbp} becomes easy to understand.
In contrast to the total quark mass $m_f + m_{\rm res}$ which depends 
significantly on both the input bare mass $m_f$ and on $L_s$ through
$m_{\rm res}$, the chiral condensate is expected to depend only on 
the input bare mass $m_f$.  In fact this dependence is quite strong
with the familiar form $m_f/a^2$.  Thus, when we keep $m_f$ fixed and
simply increase $L_s$ from 32 to 64 we should expect little change in
$\langle\overline{\psi}_q\psi_q\rangle$ as is shown in Fig.~\ref{fig:pbp}.
However, for the second set of points where $L_s$ is increased to 96
and $m_f$ is also increased to keep $m_f =m_{\rm res}$ fixed, the increase
in the bare input quark mass $m_f$ produces a significant increase in
$\langle\overline{\psi}_q\psi_q\rangle$.

As will become clear below, the above discussion of the chiral condensate
is approximate, focusing on the dominant explicit chiral symmetry breaking
term $m_f/a^2$ coming from the input quark mass and a residual chiral
symmetry breaking piece expected to behave as $\exp(-\lambda_cL_s)/a^3$.
The more interesting, physical contribution to the chiral condensate which 
arises from vacuum symmetry breaking and is described, for example, by the 
Banks-Casher formula, will depend on the physical quark mass, 
$m_f+m_\mathrm{res}$.  Such dependence on $m_\mathrm{res}$ will necessarily 
introduce a $1/L_s$ dependence on $L_s$, not seen in the results described 
in the paragraph above.  This is to be expected because the much larger 
$m_f/a^2$ and $\exp(-\lambda_cL_s)/a^3$ terms do not show this behavior. 

In contrast to the chiral condensate, the disconnected part of the chiral
susceptibility is more physical and grows with decreasing quark mass.  It 
is dominated by the large fluctuations present in the long-distance modes.  
The large $m_f/a^2$ and $\exp(-\lambda_cL_s)/a^3$ which dominate the
averaged $\langle \overline{\psi}_q\psi_q\rangle$ fluctuate less because
of the large number of short distance modes and hence contribute relatively
little to the fluctuations in the quantity $\overline{\psi}_q\psi_q$.
This behavior should be contrasted to that of the connected chiral 
susceptibility which is again dominated by short-distance modes and hence 
of less interest and not considered here. 

\begin{figure}[ht]
\begin{center}
\includegraphics[width=\textwidth]{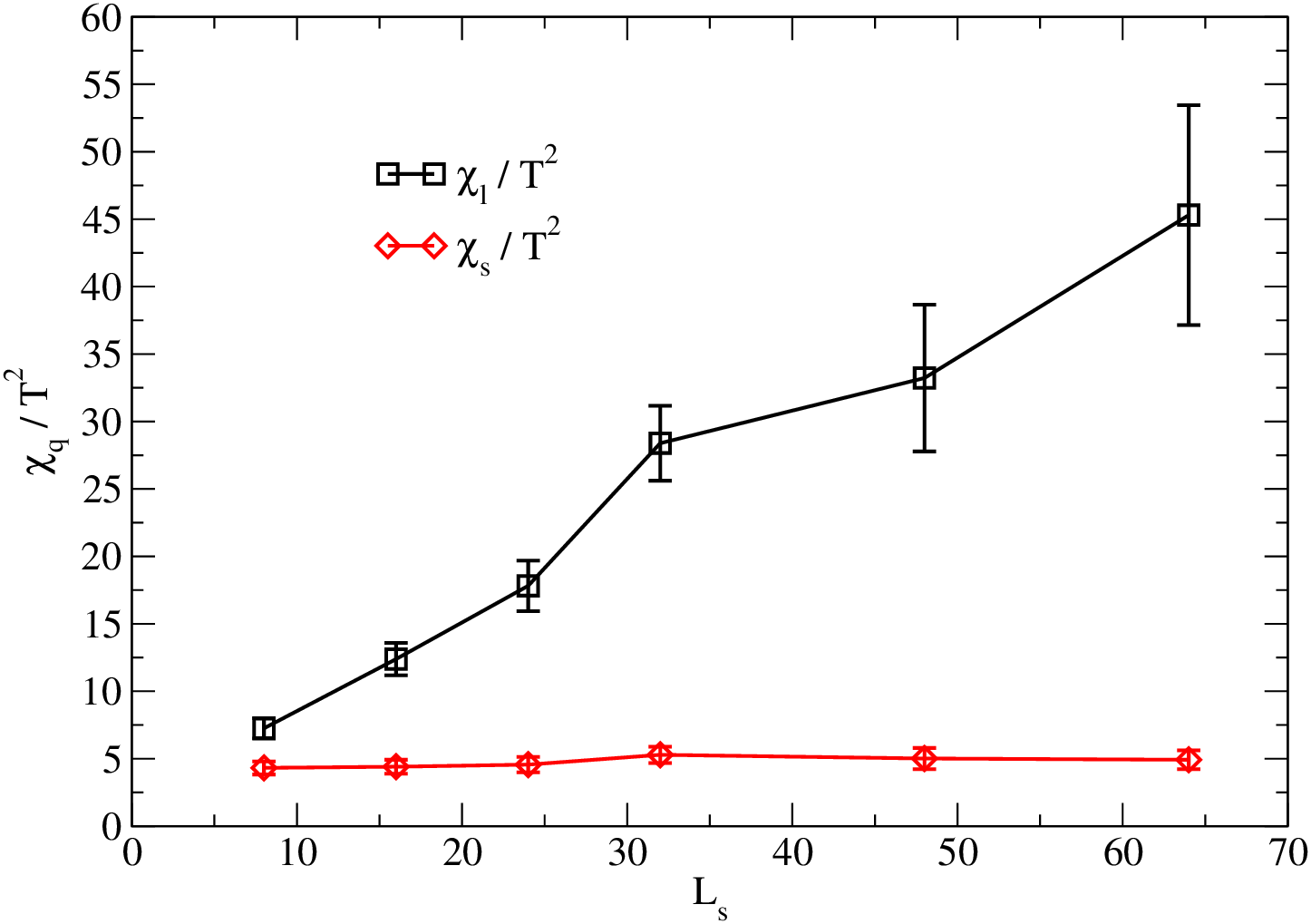}
\end{center}
\caption{Disconnected chiral susceptibility versus $L_s$ for 
$\beta = 2.0375$, $16^3\times8$, with input quark masses fixed to $m_l = 0.003$ and $m_s = 0.037$.}
\label{fig:pbp_b2.0375_sus}
\end{figure}

Thus, for small quark mass and $\beta \approx \beta_c$ we expect that the
disconnected chiral susceptibility will depend on the total effective quark mass, 
$m_q = m_l + m_\mathrm{res}$, that enters into the low energy QCD Lagrangian.
Figure \ref{fig:pbp_b2.0375_sus} shows the disconnected chiral susceptibility
at $\beta = 2.0375$ as a function of the valence $L_s$.  The chiral susceptibility
does not plateau as $L_s$ grows.  Rather, it increases as
the total quark mass $m_q = m_l + m_\mathrm{res}$ is decreased as we move to 
larger $L_s$.  The fact that the chiral susceptibility depends only on the
total quark mass can also be seen in the measurements at $L_s = 96$, where
the input quark masses are adjusted to keep the total quark mass fixed.
As we can see in Fig.~\ref{fig:pbp_sus}, the chiral susceptibility at $L_s = 96$ 
is roughly the same as at $L_s = 32$, even though the relative sizes of 
the input quark masses and the residual mass has changed dramatically.
This behavior provides a reassuring consistency check on the DWF approach:
even at finite temperature the light fermion modes carry the expected quark 
mass, $m_q = m_l + m_\mathrm{res}$.

\section{Locating $T_c$}
\label{sec:Tc}
We will now attempt to combine our finite and zero temperature results
to determine the pseudo-critical temperature, $T_c$.  As discussed in 
Section~\ref{sec:finite} and shown in Fig.~\ref{fig:pbp_sus}, the 
chiral susceptibility shows a clear peak whose location gives a value 
for $\beta_c$.  The result for $\beta_c$ is consistent with the region 
of rapid increase in the Polyakov loop and quark number susceptibilities 
seen in Figs.~\ref{fig:wline} and \ref{fig:results-2}.  Even though 
$\beta_c$ is fairly well resolved, there are still significant 
uncertainties in extracting a physical value of $T_c$ from our 
calculation.  The most important issues are:

\begin{itemize}
\item The distortion in the dependence of the chiral susceptibility 
on $\beta$ induced by the variation of $m_\mathrm{res}$ with $\beta$.
\item The uncertainty in determining the lattice scale at the peak location 
near $\beta_c = 2.0375$ from our calculation of $r_0/a$ at $\beta = 2.025$, 
performed with light quarks considerably more massive than that those 
found in nature.
\item The absence of chiral and continuum extrapolations.
\end{itemize}

We address each of these sources of uncertainty in turn.

\subsection{Correcting for $m_\mathrm{res}(\beta)$}

In Section \ref{sec:finite}, we observed that the chiral susceptibility
has a peak near $\beta = 2.0375$, which we can identify as the center
of the transition region.  However, the total light quark mass
$m_q = m_l + m_\mathrm{res}$ is different for each value of $\beta$ 
because of the changing residual mass $m_\mathrm{res}(\beta)$.  This 
changing quark mass distorts the shape of the chiral susceptibility 
curve, shifting the location of its peak from what would be seen
were we to have held the quark mass $m_q = m_l + m_\mathrm{res}$
fixed as $\beta$ was varied.

\begin{table}[hbt]
\centering
\begin{tabular}{c|cc|cc}
& \multicolumn{2}{|c|}{Gaussian} & \multicolumn{2}{|c}{Lorentz}\\
$\alpha$ & $\beta_c$ & $\chi^2$/dof & $\beta_c$ & $\chi^2$/dof\\
\hline
0 & ~2.041(2) & 1.7 & ~2.041(2) & 2.3\\
1/2 & ~2.036(3) & 1.7 & ~2.035(3) & 1.7\\
1 & ~2.030(3) & 1.7 & ~2.030(3) & 1.8\\
3/2 & ~2.024(5) & 1.8 & ~2.026(3) & 2.0\\
\end{tabular}
\caption{The corrected peak location ($\beta_c$) in the light chiral 
susceptibility determined from fits to Lorentzian and Gaussian peak 
shapes resulting from different assumptions for the light quark mass 
dependence of $\chi_l$: $\chi_l/T^2 \sim 1/(m_l + m_{res})^\alpha$).  
All fits include the 7 data points nearest the peak location, 
\textit{i.e.} $\beta \in [2.00,2.08]$.}
\label{tab:peak_location}
\end{table}

\begin{figure}[ht]
\begin{center}
\includegraphics[width=\textwidth]{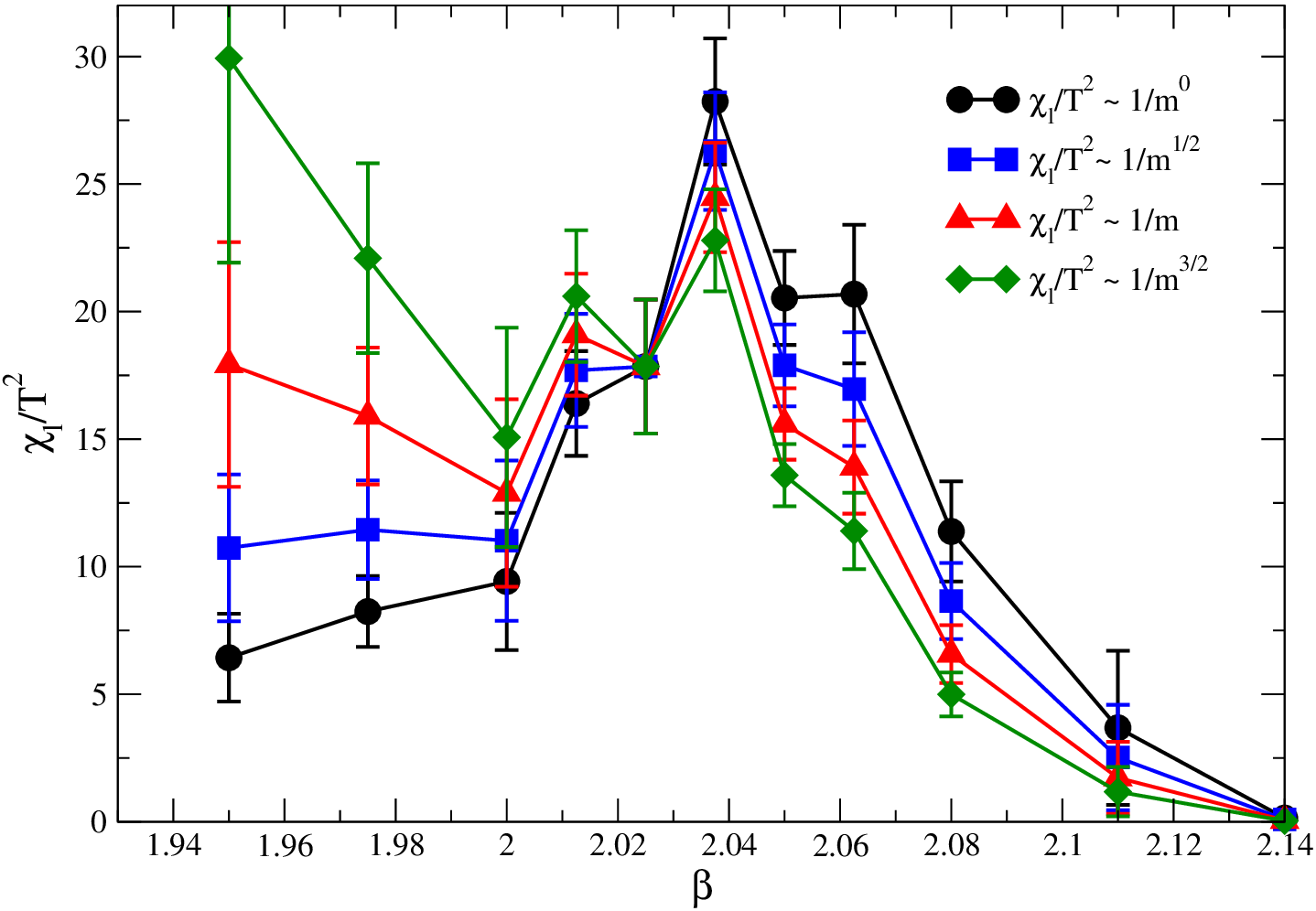}
\end{center}
\caption{Light quark chiral susceptibility, where different assumptions for 
mass dependence are used to adjust the data to a constant bare light quark 
mass $(m_l + m_{\mathrm{res}})a = 0.0097$, corresponding to the value at
$\beta = 2.025$, $L_s = 32$ in our simulations. }
\label{fig:pbp_sus_mass}
\end{figure}

In order to correct for this effect, we must account for the quark mass
dependence of the chiral susceptibility.  Our valence measurements at 
$L_s = 64$ and $L_s = 96$ indicate that the chiral susceptibility is 
inversely related to the quark mass and depends only on the combination 
$m_q = m_l + m_\mathrm{res}$.  Figure \ref{fig:pbp_sus_mass} shows the 
resulting chiral susceptibility, when one corrects for the known $\beta$
dependence of $m_\mathrm{res}(\beta)$ by assuming a power-law dependence 
of $\chi_l \propto 1/m_q^\alpha$ on the quark mass for various choices of
the power $\alpha$ ranging between $\alpha=0$ and $\alpha=3/2$.  

While for $T \le T_c$ and in the limit of small quark mass the chiral 
susceptibility is expected to behave as $\propto 1/\sqrt{m_q}$~\cite{Gasser:1986vb,
Hasenfratz:1989pk,Smilga:1993in, Smilga:1995qf,Ejiri:2009ac} corresponding 
to $\alpha = 1/2$, our data from the $L_s = 64$ valence measurements 
suggest $\alpha \sim 1.2-1.8$, albeit with rather large uncertainty.
While $\alpha > 0.5$ is inconsistent with the expected chiral behavior, 
we conservatively include such larger exponents as a possible behavior
over our limited range of non-zero quark mass.  Adjusting the chiral 
susceptibility curve in this manner enhances the chiral susceptibility at 
stronger coupling, as $m_\mathrm{res}(\beta)$ is larger on the coarser 
lattices.  This causes a systematic shift in the peak location to stronger 
coupling when this correction is made.  

While a cursory examination of Fig.~\ref{fig:pbp_sus_mass} suggests that 
this correction does not change the peak structure, more careful study 
reveals that for the extreme $\alpha=1.5$ case the peak may have 
disappeared if the two lowest $\beta$ values with large errors are taken 
seriously.  We view this possibility as unlikely but not absolutely ruled 
out.

Table \ref{tab:peak_location} gives the results of fitting the peak
region to Lorentzian and Gaussian peak shapes for various $\alpha.$
If we make no adjustment to the raw data ($\alpha = 0$),
we obtain $\beta_c = 2.041(2)$.  However, with $\alpha = 3/2$, we have
$\beta_c = 2.024(5)$ with the Gaussian fit.  While $\alpha = 3/2$ seems
to be favored by our valence measurements, we would like to emphasize that 
the quark mass dependence of the chiral susceptibility has large 
uncertainties.  In particular, since we performed valence measurements at 
only three values of $\beta$, it is unclear if this $\alpha \approx 3/2$ 
behavior holds over a broader range in $\beta$.  Also, we do not
know whether the same mass dependence will persist if both the valence
and dynamical quark masses are varied.

It should be recognized that if $\chi_l \propto 1/m_q^\alpha$
behavior for $T \le T_c$ persists in the limit of vanishing $m_q$ 
the peak structure suggested by Fig.~\ref{fig:pbp_sus_mass} may 
take on the appearance of a shoulder as the $\chi_l$ grows
for $T < T_c$.  Such a singular behavior at small quark mass,
for example the $\alpha = 1/2$ case suggested by chiral symmetry,
would make $\chi_l$ a poor observable to locate the finite 
temperature transition~\cite{Karsch:2008ch}.  Although our data 
shows an easily identified peak, unclouded by a large $1/\sqrt{m_q}$ 
term for $T \le T_c$, it is possible that such behavior may 
substantially distort the chiral susceptibility as the light 
quark mass is decreased from that studied here to its physical 
value.

With these caveats in mind, we estimate the pseudo-critical coupling 
to be $\beta_c = 2.03(1)$.  The central value corresponds to the peak 
location if we assume a quark mass dependence of 
$\chi_l \sim 1/(m_q + m_\mathrm{res})$.  The quoted error reflects 
the uncertainty in the mass dependence of $\chi_l$, and is chosen to 
encompass the range of values for $\beta_c$ shown in 
Table~\ref{tab:peak_location}.

\subsection{Extracting the lattice scale at $\beta_c$}

This value of $\beta_c$ differs from that of our zero-temperature ensemble
($\beta = 2.025$) where we have measured the Sommer parameter, $r_0/a$.
Thus, in order to determine the lattice scale at $\beta_c$, we
need to know the dependence of $r_0/a$ on $\beta$.  Fortunately,
in addition to our measurements at $\beta = 2.025$, $r_0/a$ has been
extensively measured at $\beta = 2.13$~\cite{Li:2006gra}.

At $\beta = 2.13$, the value of $r_0/a$ at the quark mass corresponding 
most closely to the current calculation is $r_0/a = 3.997(22)$.  
Extrapolation to the chiral limit gives $r_0/a = 4.113(31)$ for $\beta=2.13$, 
an approximately $3\%$ increase.  A study of finite volume effects in 
Ref.~\cite{Li:2006gra} suggests that, in addition, the value computed
on a $16^3 \times 32$ lattice is too low by approximately $1-2\%$.

To obtain $r_0/a$ at $\beta_c$, we use an exponential interpolation
in $\beta$, giving $r_0/a = 3.12(13)$, which includes the statistical
errors for $r_0/a$ and the uncertainty in $\beta_c = 2.03(1)$. 
To account for chiral extrapolation and finite volume effects, we 
add $4\%$ to this central value and also add a $4\%$ error in quadrature,
resulting in $r_0/a = 3.25(18)$.  This corresponds to $T_c r_0 = 0.406(23)$.

\subsection{Chiral and Continuum Extrapolations}

In the end, we wish to obtain a value for the pseudo-critical temperature $T_c$
corresponding to physical quark masses and in the continuum ($a \rightarrow 0$) limit.  
However, our current calculation 
is performed with a single value for the light quark masses, ($m_l/m_s \approx 0.25$), 
and a single value for the temporal extent ($N_t = 8$).  Thus, we are not 
at present able to perform a direct chiral or continuum extrapolation.

We can make an estimate of the shift in $T_c$ that might be expected when
the light quark mass is reduced to its physical value by examining the
dependence of $T_c$ on the light quark mass found in the $N_t=6$, staggered
fermion calculations in Ref.~\cite{Cheng:2006qk}.  The quark mass
dependence of $T_c$ found in Table IV of that paper, suggests a 3\% decrease
in $T_c$ when one goes to the limit of physical quark masses.

The effects of finite lattice spacing on our result can be estimated
from the scaling errors that have been found in recent zero temperature
DWF calculations~\cite{Kelly:2009xx,Mawhinney:2009xx}.  Here hadronic masses
and decay constants were studied on a physical volume of side roughly 3 fm 
using two different lattice spacings: $1/a=1.73$ and $2.32$ GeV.
The approximate 1-2\% differences seen between physically equivalent ratios 
in this work suggests fractional lattice spacing errors given by 
$(a\Lambda)^2$ where $\Lambda \approx 260-370$ MeV.  If this 
description applies as well for the $a^{-1} \approx 1.3$ GeV lattice spacing being 
used here, we expect deviations from the continuum limit of 4-7\%.

Thus,to account for the systematic uncertainty in failing to perform chiral 
and continuum extrapolations, we add a $10\%$ systematic uncertainty to
our final value for the pseudo-critical temperature, giving
$T_c r_0 = 0.406(23)(41)$.  Using $r_0/a = 0.469(7)$ fm,
this corresponds to $T_c = 171(10)(17)$ MeV.
Here the first error represents the combined statistical and systematic
error in determining $T_c r_0$ for our $a^{-1} \approx 1.3$ GeV lattice
spacing and light quark mass of $\approx 0.22$ times the strange mass.
The second error is an estimate of the systematic error associated with
this finite lattice spacing and unphysically large light quark mass.

\section{Conclusion and Outlook}
\label{sec:conclusion}

We have carried out a first study of the QCD phase transition using
chiral, domain wall quarks on a finite temperature lattice with temporal 
extent $N_t=8$.  This work represents a advance over earlier domain wall 
calculations~\cite{Chen:2000zu,Christ:2002pg} with $N_t=4$ and $6$, 
having significantly smaller residual chiral symmetry breaking and 
including important tests of the physical interpretation of the 
resulting residual mass.  Most significant is the comparison of the 
residual mass computed at fixed $\beta=2.025$ for both zero and 
finite temperature yielding $m_\mathrm{res}= 0.0069(5)$ and 0.006647(84) 
respectively.  The equality of these two results suggests that 
$m_\mathrm{res}$ can indeed be interpreted as a short-distance effect 
which acts as a small additive mass shift over the range of temperatures 
which we study.

\begin{figure}[ht]
\begin{center}
\includegraphics[width=\textwidth]{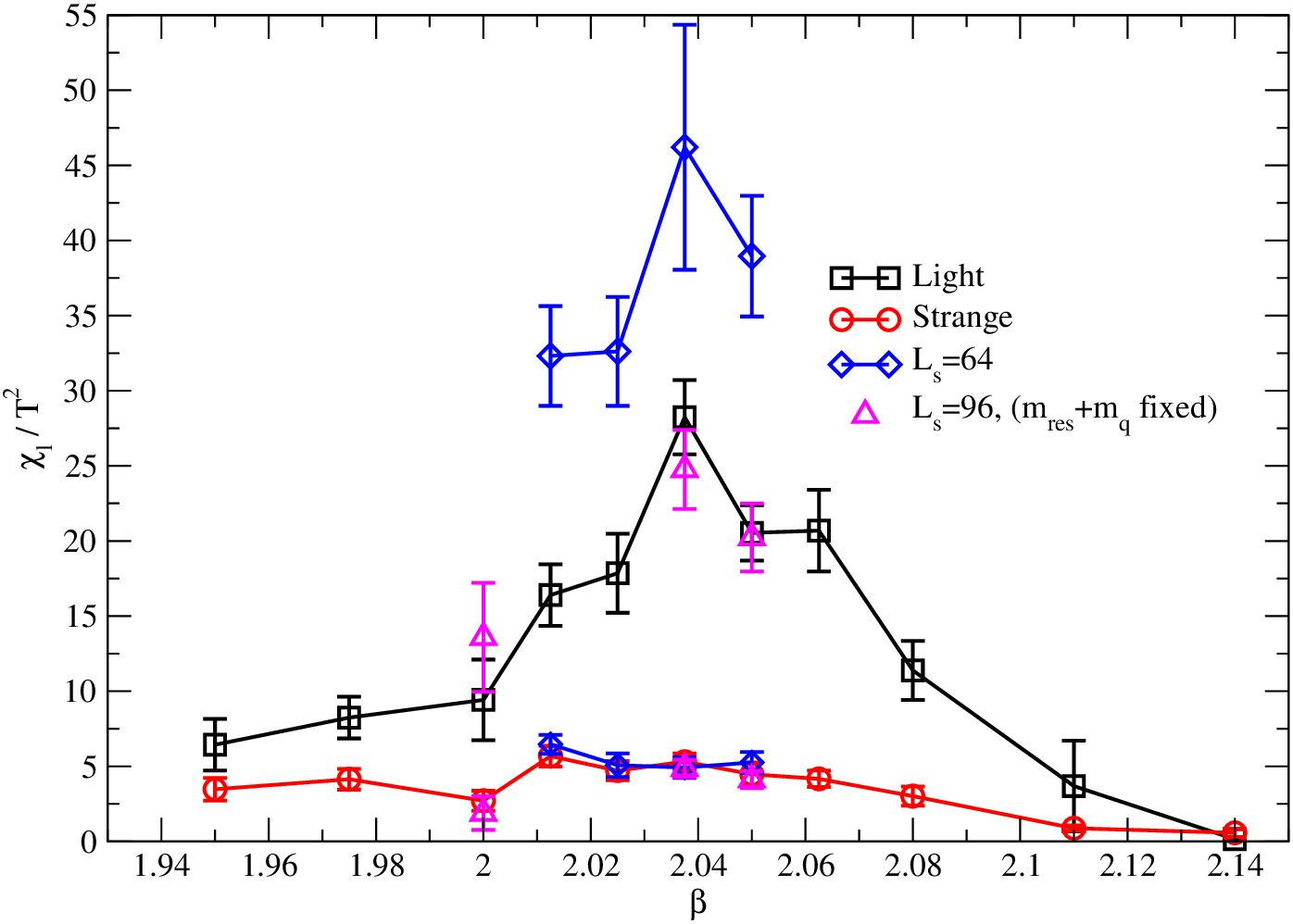}
\end{center}
\caption{Unitary values for the disconnected chiral susceptibility 
as well as the results of additional measurements with $L_s = 64$ and $L_s = 96$ 
for the valence quarks.}
\label{fig:pbp_sus}
\end{figure}

As can be seen in Fig.~\ref{fig:pbp_sus} the chiral susceptibility shows
a clear peak around $\beta_c = 2.03(1)$ and suggests a critical region 
between 155 and 185 MeV.  The peak location can be used to estimate
a pseudo-critical temperature $T_c r_0 = 0.406(23)(41)$ or 
$T_c = 171(10)(17)$ MeV.  The first error represents the statistical 
and systematic uncertainties in determining $\beta_c$ and the corresponding
physical scale at our larger than physical quark mass ($m_\pi = 308$ MeV) 
and non-zero lattice spacing, $a^{-1} \approx 1.3$ GeV.  The second error is our 
estimate of the shift that might be expected in $T_c$ as the quark mass 
is lowered to its physical value and the continuum limit is taken.  

The transition region identified from the peak in the chiral susceptibility
$\chi_l$ shown in Fig.~\ref{fig:pbp_sus} agrees nicely with the region of 
rapid rise of the Polyakov line $L$ shown in Fig.~\ref{fig:wline} and the 
charge and isospin susceptibilities, $c_2^Q$ and $c_2^I$, shown in 
Fig.~\ref{fig:results-2}.  This coincidence of the transition region 
indicated by observables related to vacuum chiral symmetry breaking ($\chi_l$)
and those sensitive to the effects of deconfinement ($L$, $c_2^Q$ and $c_2^I$)
suggests that these two phenomena are the result of a single crossover 
transition.

It is of considerable interest to compare this result with those obtained 
in two recent large-scale studies using staggered fermions 
\cite{Cheng:2006qk,Aoki:2009sc}.  Unfortunately, because of our large 
uncertainties, our result is consistent with both of these conflicting 
determinations of $T_c$.

However, there are now substantial opportunities to improve on the
calculation presented here.  Most important the size of residual chiral 
symmetry breaking must be substantially reduced.  This could be achieved
directly for the calculation described here by simply increasing the size 
of the fifth dimension.  Of course, such an increase in $L_s$ incurs 
significant computational cost.  Never-the-less, a study similar to that 
reported here is presently being carried out by the HotQCD collaboration 
using $L_s=96.$  This will provide an improved result for the chiral 
susceptibility as a function of temperature, giving a new version of 
Fig.~\ref{fig:pbp_sus} in which the total quark mass, $m_f + m_\mathrm{res}$, 
remains constant across the transition region.

More promising for large-volume domain wall fermion calculations is
the use of a modified gauge action, carefully constructed to partially
suppress the topological tunneling which induces the dominant $1/L_s$ 
term in Eq.~\ref{eqn:mres}~\cite{Vranas:1999rz,Vranas:2006zk,Fukaya:2006vs,
Renfrew:2009wu}.  This is accomplished by adding the ratio of 4-dimension 
Wilson determinants for irrelevant, negative mass fermion degrees of 
freedom to the action.  Preliminary results~\cite{Renfrew:2009wu} indicate 
that without increasing $L_s$ beyond 32, this improved gauge action 
can reduce the residual mass in the $N_t=8$ critical region by 
perhaps a factor of 5 below its current value while maintaining an 
adequate rate of topological tunneling.  This improvement, when combined 
with the next generation of computers should permit a thorough study of 
the QCD phase transition at a variety of quark masses, approaching the 
physical value and on larger physical spatial volumes.  

It is hoped that such a study of the QCD chiral transition with a fermion 
formulation that respects chiral symmetry at finite lattice spacing will 
yield an increasingly accurate quantitative description of and greater 
insight into the behavior of QCD at finite temperature.

\bibliography{dwf_thermo}

\section*{Acknowledgments}

We would like to thank Chulwoo Jung, Christian Schmidt and our other 
collaborators in the RBC-Bielefeld and HotQCD collaborations for helpful discussions.
This work has been carried out on the QCDOC computer at Columbia University
and on the computers of the New York Center for Computational Sciences 
at Stony Brook University/Brookhaven National Laboratory which is supported by 
the U.S. Department of Energy under Contract No. DE-AC02-98CH10886 and by the 
State of New York.  The work was supported in part by the U.S. Department of
Energy under grant number DE-FG02-92ER40699  and contract number DE-AC02-98CH10886.

\pagebreak
\section{Figures}


\end{document}